\documentclass[epsfig,12pt]{article}
\usepackage{epsfig}
\usepackage{graphicx}
\usepackage{hyperref}

\usepackage{array}
\usepackage{amsmath}
\usepackage{amssymb}

\newcommand{\beq}{\begin{equation}}   
\newcommand{\eeq}{\end{equation}}
\newcommand{\beqn}{\begin{eqnarray}}   
\newcommand{\eeqn}{\end{eqnarray}}

\def\be{\begin{equation}\begin{gathered}}
		\def\ee{\end{gathered}\end{equation}}

\newcommand{\diag}{\operatorname{diag}}

\newcommand{\im}{\operatorname{im}}
\newcommand{\rk}{\operatorname{rk}}

\usepackage{subcaption}
\usepackage{slashed}
\usepackage{xcolor}

\newtheorem{lemma}{Lemma}

\begin{document}

\hypersetup{%
	linkbordercolor=blue,
}

\unitlength = 1mm

\def\de{\partial}
\def\Tr{ \hbox{\rm Tr}}
\def\const{\hbox {\rm const.}}  
\def\o{\over}
\def\im{\hbox{\rm Im}}
\def\re{\hbox{\rm Re}}
\def\bra{\langle}\def\ket{\rangle}
\def\Arg{\hbox {\rm Arg}}
\def\Re{\hbox {\rm Re}}
\def\Im{\hbox {\rm Im}}
\def\diag{\hbox{\rm diag}}


\def\QATOPD#1#2#3#4{{#3 \atopwithdelims#1#2 #4}}
\def\stackunder#1#2{\mathrel{\mathop{#2}\limits_{#1}}}
\def\stackreb#1#2{\mathrel{\mathop{#2}\limits_{#1}}}
\def\Tr{{\rm Tr}}
\def\res{{\rm res}}
\def\Bf#1{\mbox{\boldmath $#1$}}
\def\balpha{{\Bf\alpha}}
\def\bbeta{{\Bf\beta}}
\def\bgamma{{\Bf\gamma}}
\def\bnu{{\Bf\nu}}
\def\bmu{{\Bf\mu}}
\def\bphi{{\Bf\phi}}
\def\bPhi{{\Bf\Phi}}
\def\bomega{{\Bf\omega}}
\def\blambda{{\Bf\lambda}}
\def\brho{{\Bf\rho}}
\def\bsigma{{\bfit\sigma}}
\def\bxi{{\Bf\xi}}
\def\bbeta{{\Bf\eta}}
\def\d{\partial}
\def\der#1#2{\frac{\d{#1}}{\d{#2}}}
\def\Im{{\rm Im}}
\def\Re{{\rm Re}}
\def\rank{{\rm rank}}
\def\diag{{\rm diag}}
\def\2{{1\over 2}}
\def\ntwo{${\mathcal N}=2\;$}
\def\nfour{${\mathcal N}=4\;$}
\def\none{${\mathcal N}=1\;$}
\def\ntwot{${\mathcal N}=(2,2)\;$}
\def\ntwoo{${\mathcal N}=(0,2)\;$}
\def\x{\stackrel{\otimes}{,}}

\newcommand{\cpn}{$\mathbb{CP}^{N-1}$}
\newcommand{\wcpn}{wCP$_{N,\widetilde{N}}(N_f-1)\;$}
\newcommand{\wcpd}{wCP$_{\widetilde{N},N}(N_f-1)\;$}
\newcommand{\wcpt}{$\mathbb{WCP}(2,2)\;$}
\newcommand{\wcpN}{$\mathbb{WCP}(N,N)\;$}
\newcommand{\wcpo}{$\mathbb{WCP}(1,1)\;$}
\newcommand{\wcp}{$\mathbb{WCP}(N,\tilde N)\;$}
\newcommand{\vp}{\varphi}
\newcommand{\pt}{\partial}
\newcommand{\tN}{\widetilde{N}}
\newcommand{\ve}{\varepsilon}
\renewcommand{\theequation}{\thesection.\arabic{equation}}

\newcommand{\sun}{SU$(N)\;$}

\setcounter{footnote}0

\vfill

\begin{titlepage}

\begin{flushright}
\end{flushright}

\begin{center}
{  \Large \bf  
   2d Sigma Models on Non-compact Calabi-Yau and ${\mathcal N}=2$ Liouville Theory
}

\vspace{5mm}

{\large  \bf Pavlo Gavrylenko$^{\,\,a}$, Evgenii Ievlev$^{\,b,c}$,\\ Andrei Marshakov$^{\,\,d,e,f}$, Ilia Monastyrskii$^{\,\,e}$ and  \\ Alexei Yung$^{\,\,b,d,e}$}
\end {center}

\begin{center}
$^{a}${\it International School for Advanced Studies (SISSA)}\\
$^{b}${\it 
	Petersburg Nuclear Physics Institute
}\\
$^{c}${\it 
	Nazarbayev University
	\& Al-Farabi Kazakh National University
}\\
$^{d}${\it Igor Krichever Center for Advanced Studies, Skoltech
}\\	
$^{e}${\it Dept. Math. \& St.Petersburg branch, HSE University
}\\
$^{f}${\it Theory Department of LPI
}\\

\end{center}

\vspace{1cm}

\begin{center}
{\large\bf Abstract}
\end{center}
We consider a class of two dimensional  conformal ${\mathcal N}=2$ supersymmetric $U(1)$ gauge linear sigma models with $N$ fields of charges $+1$ and $N$ fields of charges $-1$, whose Higgs branches are non-compact toric Calabi-Yau manifolds of complex dimension $2N-1$. We show, starting from large-$N$ approximation, that the Coulomb branch of these models, which opens up at strong coupling, is described by ${\mathcal N}=2$ Liouville theory and then extrapolate it to exact equivalence demanding the central charge of the Liouville theory to be $\hat{c}=2N-1$. Next we concentrate on mostly physically attractive $N=2$ and $N \geq 3$ cases and find there a perfect agreement of the set of complex moduli on the Calabi-Yau side with the marginal deformations in ${\mathcal N}=2$ Liouville theory, supporting proposed exact equivalence.


\setcounter{tocdepth}{2}

\newpage
\tableofcontents

\end{titlepage}



\section {Introduction }
\label{intro}
\setcounter{equation}{0}

Non-compact Calabi-Yau (CY) spaces with an isolated singularity unexpectedly emerge in the description of solitonic vortex strings in four-dimensional 
(4d) \ntwo supersymmetric QCD. 

In particular, it was shown in \cite{SYcstring} that the non-Abelian solitonic vortex string \cite{HT1,ABEKY,SYmon,HT2} 
(see \cite{Trev,Jrev,SYrev,Trev2} for reviews) in the theory with the U($N=2$) gauge group and $N_f = 2N=4$
flavors of quark hypermultiplets can be considered a critical superstring. In addition to four
translational moduli, these non-Abelian strings carry six extra (orientational and size) moduli.
Together, they form a ten-dimensional space required for a critical superstring.  
The target space of the string sigma model (in addition to $\mathbb{R}^4$) contains a non-compact CY threefold $Y_6$, which is the conifold \cite{Candel,NVafa}. The
spectrum of low-lying closed string excitations was found in \cite{KSYconifold,SYlittles}.

Most massless and massive string modes have non-normalizable wave functions over the conifold $Y_6$, i.e., they are not localized in 4d 
and cannot be interpreted as dynamical states in 4d theory. In particular, there are no massless 4d gravitons in the physical spectrum \cite{KSYconifold}.
However, an excitation associated with the deformation of the complex structure of $Y_6$ has a (logarithmically) normalizable wave function (this state is localized near the conifold singularity) and was interpreted as a baryon in the spectrum of hadrons in 4d \ntwo supersymmetric QCD (SQCD).

To analyze the massive states, it is better to use an approach 
similar to the one used for Little String Theories (LSTs) (see \cite{Kutasov} for a review), based
on the equivalence \cite{GVafa} between the 
critical string on the conifold and the non-critical $c=1$ string containing the Liouville 
field and a compact scalar  
at the self-dual radius (to be unified into a complex scalar of \ntwo Liouville theory \cite{Ivanov,KutSeib})~\footnote{In \cite{GVafa}, this equivalence was shown for topological versions of string theories.}. 
Later, a similar correspondence was proposed (and treated as a holographic AdS/CFT-type duality)
for a critical string on certain other non-compact CY spaces in the so-called double scaling limit and a non-critical $c=1$ string with an additional Landau-Ginzburg
\ntwo superconformal system \cite{GivKut,GivKutP,ES} (see also \cite{Eplus}), which is trivial in the conifold case.

The purpose of this paper is to study this equivalence in a more direct way. Namely, we aim to demonstrate that
a class of (so-called weighted $\mathbb{CP}$, where $\mathbb{CP}$ stands for the complex projective target space) $\mathbb{WCP}(N,N)$ worldsheet sigma models on non-compact toric CY manifolds with an isolated singularity 
is equivalent to the \ntwo Liouville theory. Sigma models on these CY spaces are realized as Higgs branches of $U(1)$ gauge linear sigma models (GLSMs) with $N$ fields of charge $\mathtt{Q}=+1$ and $N$ fields with charges $\mathtt{Q}=-1$. We consider arbitrary $N$, and even though the interesting cases are $N=2,3$, we sometimes apply large-$N$ arguments. The main physical motivation for studying $\mathbb{WCP}(N,N)$ models (which are conformal since $\sum_1^{2N} \mathtt{Q}=0$) comes from the observation that they emerge as worldsheet theories for non-Abelian vortex strings in 
4d \ntwo supersymmetric QCD with a U($N$) gauge group and $N_f = 2N$ quark flavors; see \cite{SYrev} for a review.


The \ntwo Liouville theory also has a mirror description \cite{HoriKapustin}, given by a supersymmetric version of Witten's two-dimensional
black hole with a semi-infinite cigar target space \cite{Wbh} or the ${\rm SL}(2,R)/{\rm U}(1)$ coset Wess-Zumino-Novikov-Witten (WZNW) conformal theory \cite{GVafa,GivKut,MukVafa,OoguriVafa95}. It can therefore be analyzed using algebraic methods of 2d CFT, and 
the spectrum of primary operators is known exactly \cite{MukVafa,DixonPeskinLy,Petrop,Hwang,EGPerry}.
These exact results were exploited in \cite{SYlittles} to obtain the low-lying spectrum of hadrons in \ntwo 4d QCD. 

The importance of the physical results obtained for the spectrum of \ntwo QCD hadrons in 4d in \cite{SYlittles} (see also 
\cite{IYcorrelators}) requires putting the above equivalence on firmer ground. 
In this paper, we show directly that the Coulomb branch of $\mathbb{WCP}(N,N)$ models
is indeed adequately described by the \ntwo Liouville theory. 
We compute, in the large-$N$ limit, the Liouville background charge to be $Q\stackreb{N\to\infty}{\approx} \sqrt{2N}$ and then argue that the exact dependence is $Q = \sqrt{2(N-1)}$. 

Next, we compare the set of complex structure moduli on the CY side with the marginal deformations 
on the Liouville side and find that, in almost all cases, both spaces are empty. The only important exception is the conifold case ($N=2$), where the only complex modulus associated with deformations of the conifold complex structure, with a (logarithmically)
normalizable wave function \cite{KSYconifold,Strom}, was found in \cite{SYlittles} to correspond to the marginal primary operator from a discrete spectrum on the Liouville side. 
We also consider in detail cases with $N\geq 3$ and show that, in these cases, CY manifolds are rigid and therefore have no complex structure moduli. This result exactly matches the absence of normalizable marginal primaries on the Liouville side. 
We also argue, from a mirror black-hole picture, that it is natural to expect the space of marginal deformations for all $N>2$ theories to be empty and relate this to the black hole/string transition.


The paper is organized as follows. In Sect.~\ref{wcpmodels}, we define \wcpN models and discuss their Higgs and Coulomb branches.
We also review the most interesting conifold case, which corresponds to the CY described by \wcpt. In Sect.~\ref{ss:Liouville},
we review the \ntwo Liouville theory, its mirror description, and the spectrum of its primary operators. In Sect.~\ref{wcp=liouville},
we show the equivalence of the Coulomb branch of the \wcpN model and the \ntwo Liouville theory, first studying the large-$N$ limit and then finding the exact formula for the Liouville background charge. We also discuss the relation
of complex structure moduli on the CY side with marginal deformations on the Liouville side. In Sect.~\ref{moduli}, we 
develop a general setup for searching for CY complex structure moduli and show that they are absent for $N\geq 3$. Sect.~\ref{conclusions} contains our conclusions. Appendix A is devoted to the derivation of Liouville interactions from the \wcp model at large $N$, while Appendix B contains explicit formulas for the $N=3$ case, illustrating 
generic considerations in Sec.~\ref{moduli}.

\section {$\mathbb{WCP}(N,N)$ sigma models on non-compact CY manifolds}
\label{wcpmodels}
\setcounter{equation}{0}

In this section we describe $\mathbb{WCP}(N,N)$ sigma models emerging as  world sheet theories for non-Abelian vortex strings in 4d \ntwo supersymmetric QCD with U($N$) gauge group and $N_f = 2N$ quark flavors.
These non-Abelian vortices are 1/2-BPS (Bogomolny-Prasad-Sommerfield) saturated therefore the world sheet theory has \ntwot supersymmetry.
First we define \wcpN  models as Higgs branches of U(1) gauge theory, and then discuss the exact twisted superpotentials known for
these models.

\subsection{$\mathbb{WCP}(N,N)$ models: Higgs branch}

The \wcpN sigma model 
can be defined as a low-energy limit of the U(1) gauge theory \cite{W93}, corresponding to the limit 
of infinite gauge coupling, $e_0^2\to\infty$. The bosonic part of this gauge linear  sigma model (GLSM) action reads
\begin{equation}
\begin{aligned}
	&S = \int d^2 x \left\{
	\left|\nabla_{\alpha} n^{i}\right|^2 
	+\left|\widetilde{\nabla}_{\alpha} \rho^j\right|^2
	-\frac1{4e_0^2}F^2_{\alpha\beta} + \frac1{e_0^2}\,
	\left|\pt_{\alpha}\sigma\right|^2 + \frac1{2e_0^2}\,D^2
	\right.
	\\[3mm]
	&-\left.
	2\left|\sigma\right|^2\left( \left|n^{i}\right|^2 +
	\left|\rho^j\right|^2\right) +D\left(\left|n^{i}\right|^2-\left|\rho^j\right|^2 - {\rm Re}\,\beta \right)
	 - \frac{\vartheta}{2\pi}F_{01}
	\right\},
	\\[4mm]
	&
	\alpha,\beta=1,...,2\,,\qquad i,j=1,...,N\,.
\end{aligned}
\label{wcpNN}
\end{equation}
where the complex scalar fields $n^{i}$ and $\rho^j$ have
charges  $\mathtt{Q}=+1$ and $\mathtt{Q}=-1$ respectively, i.e.
\begin{equation}
	\nabla_{\alpha}=\pt_{\alpha}-iA_{\alpha}\,,
	\qquad 
	\widetilde{\nabla}_{\alpha}=\pt_{\alpha}+iA_{\alpha}\,,	
	\label{cov_derivatives}
\end{equation}
 the complex scalar $\sigma$ is a superpartner of the U(1) gauge field $A_{\alpha}$ and $D$ is the auxiliary field in the vector
supermultiplet, contained in the twisted chiral superfield $\Sigma$~\footnote{Here spinor indices are written as subscripts, say 
	$\theta^L=\theta_R$, $\theta^R= -\theta_L$. We also defined the twisted measure 
	$d^2 \tilde{\theta} = \frac12\,d \bar{\theta}_{L} d\theta_{R}$ to ensure that 
	$\int d^2 \tilde{\theta}\, \tilde{\theta}^2 = \int d \bar{\theta}_{L} d\theta_{R}\, \theta_R \bar{\theta}_L=1$.} 
\beq
\Sigma = \sigma +\sqrt{2}\theta_R \bar{\lambda}_L -\sqrt{2}\bar{\theta}_L \lambda_R
+\sqrt{2}\theta_R\bar{\theta}_L (D-iF_{01})
\label{Sigma}
\eeq
with the lowest scalar component $\sigma$ \cite{W93}.
The complexified inverse coupling in \eqref{wcpNN}   
\begin{equation}
	\beta = {\rm Re}\,\beta + i \, \frac{\vartheta}{2 \pi} \,.
\label{beta_complexified}	
\end{equation}
is defined via  2d Fayet-Iliopoulos (FI) term  (twisted superpotential)
\beq
-\frac{\beta}{2}\,\int d^2 \tilde{\theta}\sqrt{2}\,\Sigma = - \frac{\beta}{2}\,(D-iF_{01})
\label{Sigma_sup}.
\eeq
It has been added to the kinetic term
\beq
S_{0}=\frac{1}{e_0^2}\,\int d^2 x d^4 \theta\bar{\Sigma}\Sigma
\label{kin_super0}
\eeq
which disappears in the limit $e_0^2\to\infty$.

The number of real  bosonic degrees of freedom in the model \eqref{wcpNN} defines the dimension of its target space (Higgs branch), given by 
\beq
{\rm dim}_\mathbb{R}{\cal H} =4N-1-1=2\,(2N-1). 
\label{dimH}
\eeq
where from $4N$ real $(n^i,\rho^j)$ fields one real $D$-term constraint 
\beq
|n^{i}|^2-|\rho^j|^2 = {\rm Re}\,\beta,
\label{D-term}
\eeq
is subtracted in the limit $e_0^2\to \infty$,  and in addition, the gauge phase is eaten by the Higgs mechanism.

At the quantum level, the coupling $\beta$ does not run in this theory because the sum  of charges of $n$ and $\rho$ fields vanishes $\sum_1^{2N} \mathtt{Q}=0$,  hence it is superconformal, at least for zero masses as in (\ref{wcpNN}). 
Therefore, its target space is Ricci flat and, being K\"ahler due to \ntwot supersymmetry, represents  a (non-compact) Calabi-Yau manifold,  see 
\cite{NVafa,Bouchard} for  reviews on toric geometry.

The dimension of the Higgs branch \eqref{dimH} determines the central charge of the 2d CFT of the 
CY manifold
\beq
\hat{c}_{CY} \equiv \frac{c_{CY}}{3} = {\rm dim}_\mathbb{C}{\cal H} = 2N-1,
\label{cCY}
\eeq
just equal to its complex dimension. 
In the $N=2$ case, these ${\rm dim}_\mathbb{R}{\cal H} = 2(2N-1)=6$ internal degrees of freedom can be combined with four translational moduli of the non-Abelian vortex to form a 10d target space of a critical superstring \cite{SYcstring,KSYconifold}, 
for the $N=3$ case ${\rm dim}_\mathbb{R}{\cal H} = 2(2N-1)=10$, so the \wcpN model itself gives rise to a critical string theory, while for $N>3$  the string theory applications of \wcpN models so far remain unclear.

The global symmetry group of the \wcpN sigma model \eqref{wcpNN} is
\beq
SU(N)\times SU(N)\times U(1)_B.
\label{global_sym}
\eeq
It is exactly the same as the unbroken global group in the
4D SQCD at $N_f =2N$ (see \cite{KSYconifold} for details), where the global $U(1)_B$ is identified with the baryonic symmetry. The fields $n^i$ and
$\rho^j$ transform in   representations 
\beq
\left(\,{\bf N}, \,{\bf 1}, \,\frac12 \right) \qquad \left(\,{\bf 1},\,{\bf N} , \,\frac12 \right)
\label{n_rho_representations}
\eeq
 respectively.  Note that another U(1) symmetry, which rotates $n$ and $\rho$ fields with the opposite charges, is
gauged in the model \eqref{wcpNN}.


\subsection{Coulomb branch}
\label{exactsup}


Classically, at strong coupling $\beta\to 0$, the D-term constraint (\ref{D-term}) allows the vanishing solution $n^i=\rho^j=0$, $\forall\ i,j$, so that the Coulomb branch with $\sigma\neq 0$ can open up in the theory (\ref{wcpNN})~\footnote{To avoid misunderstanding, let us point out that we use this terminology in a different, say, from \cite{W44} sense. Despite the general well-known problems with ill-defined branches of vacua in 2d theories due to strong fluctuations in the IR, we nevertheless refer here to their classical definitions throughout the paper, i.e., our Coulomb branch just corresponds to the sector of the theory with $\sigma\neq 0$.
\label{foot:Coulomb}}. To see how the Coulomb branch emerges in the quantum theory, we use the exact twisted superpotential for the \wcpN models obtained by integrating out $n$
and $\rho$ fields. This exact twisted superpotential is a generalization \cite{HaHo,DoHoTo}
of the CP($N-1$) model superpotential \cite{W93,AdDVecSal,ChVa,Dorey} of the Veneziano-Yankielowicz  type \cite{VYan} 
for the twisted superfield  $\Sigma$ and reads:
\begin{multline}
	 {\cal W}_{\rm WCP}(\Sigma)= -\frac{1}{4\pi}\Bigg\{ 
	 	\sum_{i=1}^{N_+} \left( \sqrt{2} \, \Sigma + m^+_i \right) \ln\left( \sqrt{2} \, \Sigma + m^+_i \right)
	 	\\
	 	- \sum_{j=1}^{N_-} \left( \sqrt{2} \, \Sigma + m^-_j \right) \ln\left( \sqrt{2} \, \Sigma + m^-_j \right)
	 	+ 2 \pi \,  \sqrt{2} \, \Sigma  \, \beta
	 	+ \text{const}
	 \Bigg\}\,,
\label{WCPsup}
\end{multline}
where we introduced twisted masses for the infrared (IR) regularization in the case of
equal numbers $N_+=N_-=N$ of positively and negatively charged multiplets. We will take the limit  $m_i^+ =m_j^-\to 0$
at the last step.

The vacuum structure of the theory (\ref*{wcpNN}) with superpotential (\ref*{WCPsup}) is given by the vacuum equation
\begin{equation}
	\prod_{i=1}^{N}\left(\sqrt{2} \, \sigma + m^+_i \right) 
		= e^{- 2 \pi \beta} \, \prod_{j = 1}^{N} \left(\sqrt{2} \, \sigma + m^-_j \right) \,
\label{2D_equation}	
\end{equation}
at generic values of parameters, giving just $N$ distinct vacua with certain fixed values of $\sigma$. In the limit 
$m^+_i=m^-_j=0$, one gets
\beq
\sigma^N = e^{- 2 \pi \beta} \,\sigma^N.
\label{sigmaeq}
\eeq
with the $N$-degenerate vacuum solution $\sigma=0$ for any nonvanishing $\beta$. This means that fields $n$ and $\rho$
remain massless and live on the Higgs branch of the theory. However, for both $\beta=0$ and vanishing twisted masses, the complex scalar $\sigma$ can have an arbitrary value, and this solution describes the Coulomb branch, which opens up at $\beta = 0$, supporting the qualitative classical picture we mentioned above. Below, we show that this Coulomb branch can be effectively described in terms of \ntwo Liouville theory.

\subsection{The resolved/deformed conifold}
\label{conifold}

As an example, we review in this section the conifold case corresponding to the \wcpN  model in  \eqref{wcpNN} with $N=2$. We shall now demonstrate that
the resolved conifold corresponds to the Higgs branch of the GLSM \eqref{wcpNN} at $N=2$, while the deformed conifold is  associated with the Coulomb branch of this theory, which opens up at $\beta=0$.

Consider the U(1) gauge-invariant ``mesonic'' variables
\beq
w^{ij}= n^i \rho^j.
\label{w}
\eeq
subject to the obvious constraint
\beq
{\rm det}\, w^{ij} =0.
\label{coni}
\eeq
This equation defines the conifold $Y_6$, and it can be endowed with
the K\"ahler Ricci-flat metric and represents, therefore, a non-compact
Calabi-Yau manifold \cite{Candel,NVafa,W93}, which can be parametrized, for example,
by the radial coordinate 
\beq
r^2 = {\rm Tr}\, \bar{w}w\,
\label{tilder}
\eeq
and five angles, so that its section at fixed $r$ is $S^2\times S^3$, see \cite{Candel}. 

At $\beta =0$, the conifold develops a conical singularity when both spheres $S^2$ and $S^3$  
shrink to zero.
The conifold singularity can be smoothed 
in two distinct ways: by deforming the K\"ahler form or by deforming the 
complex structure, both preserving the K\"ahler structure and Ricci-flatness of the metric.  

The first deformation, which amounts to keeping
a non-vanishing value of $\beta$ in (\ref{D-term}), is called the resolved conifold.
Putting $\rho^j=0$ in (\ref{wcpNN}) for $N=2$, one gets the $\mathbb{CP}^1\simeq S^2$ (with nonvanishing radius $\sqrt{\beta}$) as a 
(part of) the target space of the resolved conifold, obviously corresponding to the Higgs branch of the GLSM.
The resolved conifold has no normalizable moduli; in particular, 
its K\"ahler modulus $\beta$, becoming a scalar field for the non-Abelian string on 
$\mathbb{R}^4\times Y_6$,
has a non-normalizable (quadratically divergent) wave function over the 
conifold and therefore is not dynamical in 4d \cite{KSYconifold}.  

If $\beta=0$ (i.e. exactly when the Coulomb branch opens up), another option exists, namely a deformation 
of the complex structure \cite{NVafa},
usually referred to as the {\em deformed conifold}. 
It  is defined by the deformation of equation (\ref{coni}) 
\beq
{\rm det}\, w^{ij} = \mu\,,
\label{deformedconi}
\eeq
by a single complex parameter $\mu$, which now determines the minimal size of the 
sphere $S^3$, which can no longer shrink to zero. 

As we already mentioned, the modulus $\mu$ becomes a massless 4d complex scalar field for the non-Abelian string on 
$\mathbb{R}^4\times Y_6$. It has a logarithmically normalizable metric (with respect to the radial coordinate $r$),
which  was calculated in \cite{KSYconifold}
using the explicit metric on the deformed conifold  \cite{Candel,Ohta,KlebStrass}. This string state was interpreted in
\cite{KSYconifold} as a massless baryon of 4d SQCD.

\section {\ntwo Liouville theory}
\label{ss:Liouville}
\setcounter{equation}{0}

In this section, we briefly review the \ntwo Liouville theory, see \cite{Nakayama} for a detailed review and references therein.

\subsection{Setup}

The \ntwo Liouville theory has target space $ \mathbb{R}\times S^1_Q$,
where the real line $\mathbb{R}$ is associated with the non-compact Liouville field $\phi$, while the circle $S^1_Q$ 
corresponds to an additional compact scalar $Y \sim Y+2\pi Q$. 
The target-space background contains a dilaton linear in $\phi$
\beq
\Phi(\phi) =-\frac{Q}{2}\phi\, ,
\label{dilaton}
\eeq
so that the holomorphic stress tensor of the bosonic part of the theory is given by~\footnote{We use the normalization for the scalar fields 
	\beq
	\langle \phi(z)\phi(0)\rangle = \langle Y(z)Y(0)\rangle = -\log{z} 
	\label{propagators}
	\eeq} 
\beq
T= -\frac12\,\left[(\pt_z \phi)^2 + Q\, \pt_z^2 \phi + (\pt_z Y)^2\right]. 
\label{T--}
\eeq
The central charge of the \ntwo Liouville theory (with an additional contribution from complex fermions) is obviously
\beq
c_{L} = 3 + 3Q^2, \qquad \hat{c}_L\equiv \frac{c_L}{3} = 1 + Q^2.
\label{cL}
\eeq
The Liouville interaction corresponds to adding the superpotential
\beq
 L_{int}= \tilde{\mu}\int d^2\theta \, W\, ,
\label{liouville}
\eeq
where, in terms of the corresponding chiral superfield, with the lower component $\phi + iY$
\beq
W=e^{-\frac{\phi + iY}{Q}}
\label{WL}
\eeq
and $\tilde{\mu}$ is some complex parameter. 

The superpotential \eqref{liouville} is a marginal deformation since its conformal dimension w.r.t. the stress tensor 
\eqref{T--} is
\beq
\Delta\left(e^{-\frac{\phi + iY}{Q}}\right)= \frac{1}{2} \left(-\frac1{Q^2} +1 + \frac1{Q^2}\right) =\frac12
\label{DeltaLW}
\eeq
or the total (left and right) conformal dimension of the 
operator $W$ is $\left(\frac12,\frac12\right)$, i.e. exactly what is necessary to be a marginal deformation in (\ref{liouville}) after integrating over the Grassmann coordinates.


If we consider the Liouville theory as a worldsheet sigma model in string theory, then the string coupling
\beq 
g_s= e^{\Phi}= \exp{(-\frac{Q}{2} \phi)}
\label{strcoupling}
\eeq
depends on $\phi$, see \eqref{dilaton}. It goes to zero at $\phi\to\infty$, while at $\phi\to -\infty$ it becomes infinite.
At $\tilde{\mu}\neq 0$, the Liouville interaction regularizes this behavior
of the string coupling, preventing the string from propagating to the region of large negative $\phi$.

The mirror description of the \ntwo Liouville theory \cite{HoriKapustin} can be given in terms of (a supersymmetric version of) the two-dimensional
black hole \cite{Wbh}, which is the SL($2, \mathbb{R}$)/U(1) coset WZNW theory 
\cite{GVafa,GivKut,MukVafa,OoguriVafa95} 
with the level
\begin{equation}
	k =\frac{2}{Q^2}\,.
\label{k_Q_relation}
\end{equation}
of the supersymmetric version of the Ka\v{c}-Moody algebra~\footnote{The level
of the bosonic part of the algebra is $k_b=k+2$. }.
The bosonic action for the target-space metric of this theory reads
\beq
S_{\rm BH} = \frac{k}{4\pi}\int d^2 x \left\{(\pt_{\alpha} \phi_c)^2 +\tanh^2{\phi_c} \, (\pt_{\alpha}\vartheta)^2 \right\}
\label{BH}
\eeq
with the dilaton given by
\beq
\Phi(\phi_c)= \Phi_0 -\log{\cosh{\phi_c}}.
\label{dilaton_cigar}
\eeq 
with $\Phi_0\sim -\log \tilde{\mu}$, so that the target 
space has the form of a semi-infinite cigar with radial coordinate $\phi_c$ and angular coordinate $\vartheta\sim\vartheta+2\pi$.  
At $\phi_c\to\infty$, the cigar becomes just a cylinder with the radius $\sqrt{2k}$, dual to the radius $Q=\sqrt{2/k}$ of the cylinder of Liouville theory. 
As in Liouville theory, one gets here a semi-infinite geometry since $\phi_c$ is naturally restricted to a half-line (with the radius of the cigar shrinking to zero at  $\phi_c\to 0$), reproducing the effect of the ``Liouville wall''.

\subsection{Primary operators }
\label{sec:vertices}

The primary operators in  \ntwo Liouville theory (\cite{GivKut}, see also 
\cite{GivKutP,MukVafa}) for large 
positive $\phi$, when the Liouville interaction is small, take the free-field form
\beq
V_{j;m_L,m_R} \simeq e^{Q\left[j\phi + i(m_L Y_L -m_R Y_R)\right]},
\label{vertex}
\eeq
where $Y_{L,R}$ correspond to the left- and right-moving parts of the compact scalar, with quantum numbers $m_{L,R}$
\beq
m_L= \frac12(n_1+kn_2), \qquad  m_R= \frac12(n_1-kn_2),
\label{m}
\eeq
related to integer momentum and winding numbers $n_2$ and $n_1$, respectively (and vice versa in the mirror cigar picture).

The primary operator \eqref{vertex} is related to the corresponding target-space wave 
function $V(\phi,Y) = g_s(\phi) \Psi(\phi,Y)$ 
by the $\phi$-dependent string coupling \eqref{dilaton}, thus
\beq
\Psi_{j;m_L,m_R}(\phi,Y)\stackreb{\phi\to\infty}{\sim} e^{Q(j+\frac{1}{2})\phi + i Q (m_L Y_L - m_R Y_R)}\,.
\label{psifun}
\eeq
i.e. the states with normalizable wave functions correspond to 
\beq
j\le -\frac12\,.
\label{normalizable}
\eeq
where the borderline case $j=-\frac12$ is included. 

The conformal dimension of the operator \eqref{vertex} is 
\beq
\Delta_{j,m} = \frac{Q^2}{2}\left\{m^2 - j(j+1)\right\} = \frac1{k}\left\{m^2 - j(j+1)\right\}\, .
\label{dimV}
\eeq
Unitarity requires
\beq
\Delta_{j,m}> 0\,.
\label{Deltapositive}
\eeq
and for our string applications, these operators should obey $m_R=\pm m_L$~\footnote{In type IIA string $m_{R}= - m_{L}$, 
while for type IIB string $m_{R}= m_{L}$ \cite{SYlittmult}.}.

The spectrum of the allowed values of $j$ and $m$ in \eqref{vertex} was exactly determined using the
mirror description of the theory as a ${\rm SL}(2,R)/{\rm U}(1)$ coset in \cite{MukVafa,DixonPeskinLy,Petrop,Hwang,EGPerry}, 
see \cite{EGPerry-rev} for a review. Parameters $j$
and $m$ are then identified with the global quadratic Casimir and the spin projection
\beq
J^2\, |j,m\rangle\, = -j(j+1)\,|j,m\rangle, \qquad J_3\,|j,m\rangle\, =m \,|j,m\rangle
\eeq
and for the allowed values we have:
\begin{itemize}
	\item {\em Discrete representations} with
	\beq
	j=-\frac12, -1, -\frac32,..., \qquad m=\pm\{j, j-1,j-2,...\}.
	\label{discrete}
	\eeq
	\item {\em Principal} continuous representations with
	\beq
	j=-\frac12 +is, \qquad m= {\rm integer} \quad {\rm or} \quad m= \mbox{ half-integer},
	\label{principal}
	\eeq
	where $s$ is a real parameter.
\end{itemize}
The discrete representations include the normalizable and borderline $j=-\frac12$ states localized near the tip of the cigar, which nicely matches qualitative expectations.
For generic $j<-1/2$, not belonging to the discrete spectrum, the primary operator has the form 
\cite{Teschner:1999ug,LSZinLST,IYcorrelators}~\footnote{These formulas are commonly written using the dual cigar variables. In the weak-coupling domain, far from the tip of the cigar, they are related to variables in \eqref{BH} via $\phi_c=\frac{Q}{2}\phi$, $\vartheta_{L,R}=\pm\frac{Q}{2}Y_{L,R}$.
}
\begin{equation}
	V_{j, m_L, m_R}
	\simeq
	e^{i Q(m_L Y_L - m_R  Y_R)}\left[e^{Q j \phi}+ R(j, m_{L,R} ; k) e^{-Q(j+1) \phi}\right]
\label{genericj}
\end{equation}
and always contains an extra exponent (with dual $j'=-j-1>-1/2$ of the same conformal dimension \eqref{dimV}), giving a rising contribution at $\phi\to\infty$ to the wave function. Therefore, these primary operators are 
non-normalizable at generic values of $j$. The so-called reflection coefficient in \eqref{genericj}, given by \cite{FatB,LSZinLST}
\begin{multline}
	R(j, m_{L,R} ; k) \\
	=
		\left[ \frac{1}{\pi} \frac{\Gamma\left(1+\frac{1}{k}\right)}{\Gamma\left(1-\frac{1}{k}\right)} \right]^{2j+1} 
		\frac{\Gamma\left(1-\frac{2 j+1}{k}\right)  \Gamma(-2 j-1)}
		{\Gamma\left(1+\frac{2 j+1}{k}\right)\Gamma(2 j+1)}\prod_{m=m_L,m_R}\frac{\Gamma(m+j+1)}{ \Gamma(m-j)  }
\label{refl_GK}
\end{multline}
vanishes for values of $j$ and $m_L$, $m_R$ from the
discrete spectrum in \eqref{discrete}, and that kills the rising exponential in 
\eqref{genericj}~\footnote{For ${\rm Re}\,j =-1/2$ both exponentials are present in \eqref{genericj}, but they have
the same normalization properties.}, so that the primary operator gives a normalizable wave function, see \cite{IYcorrelators} for details.
The above representations contain
states with negative norm; to exclude them, one has to impose an extra restriction
\cite{DixonPeskinLy,Petrop,Hwang,EGPerry,EGPerry-rev} 
\beq
-\frac{k+2}{2}< j <0\,.
\label{no_ghosts}
\eeq

\section {The equivalence}
\label{wcp=liouville}
\setcounter{equation}{0}

Now we show that the Coulomb branch of the \wcpN model, which opens up at $\beta=0$, can be described in terms
of \ntwo Liouville theory. 

\subsection{Large $N$ calculation}

Consider, first, the \wcpN model \eqref{wcpNN} in the large $N\to\infty$ limit. As we discussed in Sect.~\ref{exactsup}, at $\beta=0$ the complex scalar 
$\sigma$ can take arbitrary values on the Coulomb branch of the theory.
For $\sigma\neq 0$, this makes the fields $n$ and $\rho$ massive, and one can integrate them out. Such a computation in the large $N$ approximation 
has been done in \cite{W79} for (both non-supersymmetric and \ntwot supersymmetric) \cpn models, see also \cite{SYhetN}. 
The bare gauge coupling $e_0^2$, taken to be infinite
 in the classical limit, is renormalized at one loop and becomes finite, so that the auxiliary U(1) gauge field in the GLSM formulation acquires a finite kinetic term and becomes dynamical.

\begin{figure}[h]
	\centering
	\includegraphics[width=0.5\textwidth]{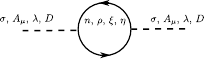}
	\caption{The wave function renormalization for the gauge multiplet.}
\label{fig:esigma}
\end{figure}

Almost the same calculation for the \wcpN model gives the effective action for the vector multiplet~\footnote{Note that the chiral anomaly does not arise in the conformal theory at hand, therefore the term ${\rm Arg}(\sigma)\,F_{01}$ is absent in 
\eqref{eff_action_kin},
 \cite{W79,SYhetN}.}  
\beq
S_{\rm eff}^{\rm kin} = \int d^2 x \left\{ -\frac1{4e^2}F^2_{\alpha\beta} + \frac1{e^2}|\pt_{\alpha}\sigma|^2 + \frac1{2e^2} D^2\right\}
\label{eff_action_kin},
\eeq
where we presented the kinetic terms of the bosonic components of the twisted superfield $\Sigma$. 
The classical gauge coupling $e^2_0$ is corrected by one loop contribution
\beq
\frac1{e^2} = \left.\left( \frac1{e_0^2} + \frac{2N}{4\pi}\,\frac{1}{2|\sigma|^2}\right)\right|_{e^2_0\to\infty} = \frac{2N}{4\pi}\,\frac{1}{2|\sigma|^2}\,.
\label{e}
\eeq
The wave function renormalization (e.g. for $\sigma$, see  Fig.~\ref{fig:esigma}) comes from $n$ and $\rho$ (with their fermionic superpartners $\xi_n$ and $\xi_{\rho}$)  propagating  in the loop. 
The loop integral is UV-finite and is saturated in the IR region at momenta of order of $n$ and $\rho$ ``mass''
$\sqrt{2}|\sigma|$, see \eqref{wcpNN}~\footnote{We put here ``mass'' in quotation marks, since in the 2d theory $\sigma$ does not have a definite vacuum expectation value (VEV), instead the ground state wave function is spread over the whole Coulomb branch, cf. \cite{W44}. See also footnote \ref{foot:Coulomb}.
\label{foot:mass}}.
The graph on Fig.~\ref{fig:esigma} contains two vertices, each proportional to the electric charge of a given $n$ or $\rho$ field (equal to $\mathtt{Q}=\pm1$), all giving rise to the coefficient $\sum_1^{2N} \mathtt{Q}^2=2N$.
The result \eqref{e} gives the leading term in the $1/N$ expansion.

The U(1) gauge field has no physical degrees of freedom in two dimensions and can be integrated out together with the $D$-field,
so we are left with the effective action for the $\sigma$ field
\beq
S_{\rm eff}^{\sigma}=\frac{2N}{4\pi}\int d^2 x \; \frac1{2}\,\frac{|\pt_{\alpha}\sigma|^2}{|\sigma|^2} + \ldots
\label{tube}
\eeq
with the tube metric~\footnote{The metric looks singular, but actually 
there is no singularity at the origin  \cite{W44}.}. 
Making a change of variables
\beq
\sigma= \gamma \,e^{-\frac{\phi + iY}{Q}},
\label{sigma}
\eeq
where $\gamma$ is a constant, to be specified below, we parametrize the modulus of $\sigma$ by  the real scalar field $\phi$ and its phase by the real compact scalar $Y$ with the periodicity condition
\beq
Y+2\pi Q \sim Y
\eeq
we arrive at the bosonic part of the effective action (for the standard normalization of kinetic terms of $\phi$ and $Y$, see \eqref{propagators})
\beq
S_{\rm eff}^{\sigma}=\frac{1}{4\pi}\int d^2 x \;\left( \frac1{2}\,(\pt_{\alpha}\phi)^2 + \frac1{2}\,(\pt_{\alpha}Y)^2\right)  + \ldots
\label{free_action}
\eeq
with the radius of the compact dimension 
\beq
Q\stackreb{N\to\infty}{\approx}\sqrt{2N}.
\label{Q_largeN}
\eeq

By this calculation up to now, we have obtained just a free field theory \eqref{free_action}, however, in order to demonstrate the equivalence of 
our effective theory  on the Coulomb branch to the \ntwo Liouville theory, we have to restore the value of the background charge for
the field $\phi$ as well as the Liouville interaction. Let us consider the background charge first.

To do this, we just repeat the above calculation on a world sheet with a nontrivial metric $ds^2=h_{\alpha\beta}dx^\alpha dx^\beta$, cf. with \cite{GMMOS}. 
The terms in the action \eqref{wcpNN} relevant for this
calculation take the form
\beq
\int d^2 x \sqrt{h}\left(
	h^{\alpha\beta}\left(\pt_{\alpha} \bar{n}_{i}\pt_{\beta} n^{i}
	+\pt_{\alpha} \bar{\rho}_{j}\pt_{\beta} \rho^{j}\right)
	+2\left|\sigma\right|^2\left( \left|n^{i}\right|^2 +
	\left|\rho^j\right|^2\right) \right),
	\eeq
where $h={\rm det}h_{\alpha\beta}$. 
Integrating by parts and using the conformal gauge
\beqn
&& \int d^2 x \left\{ \bar{n}_i \left(-\pt_{\alpha}\sqrt{h}h^{\alpha\beta}\pt_{\beta} 
+2\left|\sigma\right|^2 \sqrt{h}\right)n^{i} + \bar{\rho}_j \left(-\pt_{\alpha}\sqrt{h}h^{\alpha\beta}\pt_{\beta} 
+2\left|\sigma\right|^2 \sqrt{h}\right)\rho^{j} \right\}
\nonumber \\ 
 && 
=\int d^2 x \left\{ \bar{n}_i \left(-\pt_{\alpha}^2 
+2\left|\sigma\right|^2 \sqrt{h}\right)n^{i} + \bar{\rho}_j \left(-\pt_{\alpha}^2
+2\left|\sigma\right|^2 \sqrt{h}\right)\rho^{j} \right\},
\eeqn 
it is easy to see that the only modification
of the  calculation of diagrams, leading to the effective action in \eqref{tube},
comes from the replacements
\beq
\sigma \to \sigma (h)^{1/4}, \qquad \bar{\sigma} \to \bar{\sigma} (h)^{1/4}.
\eeq
Explicitly, instead of \eqref{tube}, one now gets for the $\sigma$ kinetic term
\beq
S_{\rm eff}^{\sigma}
= \frac{1}{4\pi}\int d^2 x\sqrt{h} \; \frac{Q^2}{2}\,h^{\alpha\beta}\pt_{\alpha}\log{(\sigma(h)^{1/4})}\pt_{\beta}\log{(\overline{\sigma}(h)^{1/4})}
\label{tube_h} 
\eeq
and dropping the $\sigma$-independent terms (the conformal anomaly vanishes for the critical string) we see that the only modification comes from the  cross-term 
$\pt_{\alpha}(\log{|\sigma|^2})\, \pt^{\alpha}\log{h}$.
Integrating again by parts and substituting (\ref{sigma})  we finally get for (\ref{tube}) on a generic world-sheet
\beq
S_{\rm eff}^{\sigma}=\frac{1}{4\pi}\int d^2 x\sqrt{h}
 \;\left(\frac1{2}\,h^{\alpha\beta}(\pt_{\alpha}\phi\pt_{\beta}\phi  +\pt_{\alpha}Y\pt_{\beta}Y)
 -\frac{Q}{2}\phi\, R^{(2)} \right),
\label{sigma_kin_R}
\eeq
where we used \eqref{sigma} and the expression for the 2d Ricci scalar in the conformal gauge 
$ R^{(2)}=-\frac1{\sqrt{h}}\pt_{\alpha}^2\log{\sqrt{h}}$.

Eq.\eqref{sigma_kin_R} is already exactly the bosonic part of the  \ntwo Liouville action leading to the energy-momentum tensor 
\eqref{T--} up to the Liouville interaction terms,
which we will consider in the next subsection. Note that the linear dilaton term has the  background charge $Q$ for the field 
$\phi$  which coincides with the radius of the compact dimension (as it should in the \ntwo Liouville theory). In the large $N$ approximation $Q$ is given by \eqref{Q_largeN}.

\subsection{Liouville interactions}

Let us now restore all other terms in the effective action for the vector supermultiplet given by one-loop calculation at large 
$N$. The effective action (on a flat world-sheet) takes the form
\be
S_{\rm eff}=\frac1{4\pi}\frac{Q^2}{2}\,\int d^2 x \left\{ \frac{|\pt_{\alpha}\sigma|^2}{|\sigma|^2} 
+ \frac{\bar{\lambda}_L}{\bar{\sigma}}\, i\pt_R \left(\frac{\lambda_L}{\sigma}\right)
+ \frac{\bar{\lambda}_R}{\bar{\sigma}}\, i\pt_L \left(\frac{\lambda_R}{\sigma}\right)\right.
\\
\left. +\frac{1}{|\sigma|^2}\left(\frac{D+iF_{01}}{\sqrt{2}} -\frac{\lambda_L\bar{\lambda}_R}{\bar{\sigma}}\right)      
\left(\frac{D-iF_{01}}{\sqrt{2}} -\frac{\lambda_R\bar{\lambda}_L}{\sigma}\right)\right\},
\label{eff_kin}
\ee
where the kinetic terms for other components of the vector multiplet were calculated in 
\cite{W79,SYhetN}, while $\lambda_{L,R}$ are fermion superpartners of the gauge field. In particular, the kinetic terms for the gauge and $D$-fields 
 \beq
-\frac{Q^2}{8|\sigma|^2}F^2_{\alpha\beta} + \frac{Q^2}{4|\sigma|^2} D^2 = \frac{Q^2}{4|\sigma|^2}(D-iF_{01})(D+iF_{01})
\eeq
in the second line of \eqref{eff_kin} are completed by the cross-terms $D\bar{\lambda}\lambda$ and $F_{01}\bar{\lambda}\lambda$ and four-fermion interaction, calculated below in Appendix~\ref{sec:loops} \footnote{There could be certainly higher derivative corrections to the  action \eqref{eff_kin}, cf. \cite{DAdda}, which flow to zero in IR.}.

In terms of twisted chiral superfields (\ref{Sigma}) the effective action \eqref{eff_kin} can be written in the form
\beq
S_{\rm eff}=\frac1{4\pi}\frac{Q^2}{2}\,\int d^2 x d^4 \theta \ln{\bar{\Sigma}}\ln{\Sigma}.
\label{eff_kin_super}
\eeq
i.e. now the K\"ahler potential is completely different from that of the original sigma-model.

As we already integrated (at $\beta=0$) the $n$ and $\rho$ fields out, and consider now the Coulomb branch of the \wcpN model, one can again switch on the 2d FI term \eqref{Sigma_sup}
\beq
S_{\rm FI}=\frac{\tilde{\mu}}{\gamma}\,\int d^2 x d^2 \tilde{\theta}\, \Sigma + c.c.= \frac{\tilde{\mu}}{\gamma}\,\frac{D-iF_{01}}{\sqrt{2}} +c.c.
\label{twist_sup}
\eeq
Now, however, when added to the kinetic term, defined by the K\"ahler potential (\ref{eff_kin_super}), this term has a different physical interpretation, 
and we show below that it reproduces the interaction induced by the Liouville superpotential \eqref{liouville}~\footnote{The fact that the Liouville interaction is given by a twisted superpotential  is just a matter of conventions since there are no untwisted chiral fields in the effective theory.}. In order to make contact between \eqref{Sigma_sup} and \eqref{liouville},
the coefficient in front of this superpotential should be taken to be  
$\tilde{\mu} =- \beta \, \gamma\,/\sqrt{2}$. Note, however, that the coefficient $\tilde{\mu}$ can remain nonvanishing even in the limit $\beta\to 0$, if 
we allow the singular behavior of $\gamma\stackreb{\beta\to 0}{\sim} 1/\beta$. 
We assume  this and discuss the 
relation of $\tilde{\mu}$ to parameters of the original \wcpN model below in Sect.~\ref{ss:exact}. It is also important to point out that the constant \(\gamma\) and parameter \(\tilde{\mu} \) are charged with respect to \(U(1)_B\), while the parameter \(\beta\) and the field \(\sigma\) (a superpartner of the gauge field) are neutral. This is important to remember in the context of the general discussion of the global symmetries.

The FI term \eqref{liouville} is a marginal deformation of the original \wcpN model, so we expect that the twisted superpotential 
\eqref{twist_sup} is also a marginal deformation of \ntwo Liouville theory. To confirm this, note that the 
complex scalar $\sigma$ is a superpartner of the gauge potential $A_{\alpha}$ (and can be constructed from extra components 
of the gauge field upon dimensional reduction, see \cite{W93}); therefore, it should have a scaling dimension equal to unity, i.e. different from the standard dimension of a scalar field in 2d. For conformal dimensions, one therefore gets 
\beq
\Delta (\sigma)= \left(\frac12,\,\frac12\right).
\label{dim_sigma}
\eeq
and this can be checked explicitly, using representation \eqref{sigma} and \eqref{DeltaLW}.

Adding the superpotential \eqref{twist_sup} to the action \eqref{eff_kin}, one gets
\beqn
&& S_{\rm eff}=\frac{1}{4\pi}\int d^2 x \;\left\{ \frac{Q^2}{2}\frac{|\pt_{\alpha}\sigma|^2}{|\sigma|^2}
 +\bar{\psi}_L i\pt_R \psi_L +\bar{\psi}_R i\pt_L \psi_R +\frac{Q^2}{2|\sigma|^2}\bar{F} F
\right.
\nonumber\\
&&
\left.
 + 4\pi\, \frac{\tilde{\mu}}{\gamma}F + 4\pi\, \frac{\bar{\tilde{\mu}}}{\bar{\gamma}}\bar{F}  + \frac{8\pi}{Q^2}\, \frac{\tilde{\mu}}{\gamma}\,\sigma\, \psi_R\bar{\psi}_L 
+\frac{8\pi}{Q^2}\, \frac{\bar{\tilde{\mu}}}{\bar{\gamma}}\,\bar{\sigma}\, \psi_L\bar{\psi}_R
\right\},
\eeqn
where instead of real variables $D$ and $F_{01}$, we introduced complex variables
\beq
F=\frac{D-iF_{01}}{\sqrt{2}} -\frac{\lambda_R\bar{\lambda}_L}{\sigma},\qquad 
\bar{F}=\frac{D+iF_{01}}{\sqrt{2}} -\frac{\lambda_L\bar{\lambda}_R}{\bar{\sigma}}
\eeq
and defined new fermionic fields,
\beq
\psi_{R}= \frac{Q}{\sqrt{2}\sigma}\, \lambda_{R}, \quad \bar{\psi}_{L}= \frac{Q}{\sqrt{2}\sigma}\, \bar{\lambda}_{L},
\quad \bar{\psi}_{R}= \frac{Q}{\sqrt{2}\bar{\sigma}}\, \bar{\lambda}_{R}, 
\quad \psi_{L}= \frac{Q}{\sqrt{2}\bar{\sigma}} \lambda_{L}.
\label{psi}
\eeq
Integrating out $F$ and $\bar{F}$, we finally get
\beqn
&& S_{\rm eff} = \int d^2 x \,\left\{ \frac{1}{4\pi}\left[ \frac1{2}\,(\pt_{\alpha}\phi)^2 + \frac1{2}\,(\pt_{\alpha}Y)^2 
-\frac{Q}{2} \phi\,R^{(2)} +\bar{\psi}_L i\pt_R \psi_L +\bar{\psi}_R i\pt_L \psi_R\right]
\right.
\nonumber\\
&& 
 +\frac{2\tilde{\mu}}{Q^2}\, \psi_R\bar{\psi}_L \,e^{-\frac{\phi + iY}{Q}}
+\frac{2\bar{\tilde{\mu}}}{Q^2}\, \psi_L\bar{\psi}_R\,e^{-\frac{\phi - iY}{Q}}
\left.
-4\pi\frac{|\tilde{\mu}|^2}{Q^2} :e^{-\frac{\phi - iY}{Q}}::e^{-\frac{\phi + iY}{Q}}:\right\}
\label{eff_action}
\eeqn
which is the action of the \ntwo Liouville theory with interaction terms induced by the superpotential \eqref{liouville}  (up to twisting) and restored linear dilaton background, see  \cite{Nakayama} for a review. The fact that the polynomial Lagrangian theory flows to exponential in the context of 2d SCFT's is actually known already for a long time, see e.g. \cite{MM90}.

\subsection{The exact equivalence
\label{ss:exact}}

So far our derivation of the \ntwo Liouville theory from the \wcpN model was based on large $N$ computation. 
Now we will argue that the equivalence of the Coulomb branch of the \wcpN model and the \ntwo Liouville theory is valid beyond the large $N$ approximation, and can be actually \emph{exact} for the corrected dependence of the Liouville background charge $Q=Q(N)$ on $N$, so that $Q^2(N)\stackreb{N\to\infty}{\approx} 2N$ at leading order. 

Suppose, indeed, that the fields $n$ and $\rho$ of the \wcpN model \eqref{wcpNN} at $\beta=0$ have been integrated out exactly, rather than in the large $N$ approximation. The $\sigma$-dependence of the effective action is actually fixed on dimensional grounds, hence we arrive at the same results as in \eqref{tube}, 
i.e., to the same conformal \ntwo Liouville theory \eqref{eff_action},
with the coefficient $2N$ replaced by some exact coefficient $Q^2(N)\stackreb{N\to\infty}{\approx} 2N$.
To fix this parameter exactly, we just require that the central charge \eqref{cL} should coincide with the central charge of the original conformal \wcpN model \eqref{wcpNN},
given by \eqref{cCY}.

Their equality 
\beq
\hat{c}_{CY} = 2N-1 = 1+Q^2 = \hat{c}_L,
\eeq
immediately gives rise to the exact relation
\beq
Q = \sqrt{2(N-1)}, \qquad k=\frac1{N-1}.
\label{Qexact}
\eeq
clearly reproducing, but correcting the previous large-$N$ calculation.

In principle, the Higgs and the Coulomb branches of 2d conformal theories can have different central charges. This was discussed in detail by Witten in \cite{W44} for the case of ${\mathcal N}=(4,4)\;$ theories. 
In particular, the central charge $\hat{c}$ on the Coulomb branch, different from the Higgs branch one, can be given by the rank of the gauge group. 
In our case, this option would give  $\hat{c} =1$, and in such a case, the conformal dimension of field $\sigma$ equals zero. However, quite similar to the reasoning from
\cite{W44} for the tube metric \eqref{tube} (as well as for the tube metric discussed in \cite{W44} for the $U(1)$ theory), $\sigma$ should rather have conformal dimension one, since it is a superpartner of the gauge potential, and the central charge then equals the dimension of the Higgs branch.

Our conjecture follows the same logic. Moreover, the option with  $\hat{c} =1$ contradicts our large $N$ calculation, which shows that the nonvanishing background charge of $\phi$ is induced in the Liouville world sheet theory \eqref{sigma_kin_R} and ensures that $\hat{c} \approx 2N$ for large $N$. This background charge leads to the nonzero conformal dimension of $\sigma$ (equal to unity). All this leads to the conclusion that  $\hat{c}=2N-1$ coincides with the dimension of the Higgs branch. This is also confirmed by more delicate arguments, such as the Coulomb and Higgs branches not being really distinguished in 2d due to strong IR effects. Moreover, in our theory, the Coulomb branch is not actually present in quantum theory except for the \(N=2\) case, see below.

The next natural question is about the interpretation of the new  parameter $\tilde{\mu}$ in \eqref{twist_sup} and \eqref{eff_action}. This is the coefficient in front of the only holomorphic marginal deformation of the \ntwo Liouville theory, given by the twisted superpotential, which suggests that CY manifolds, described by \wcpN models, may have corresponding deformations, preserving the Ricci-flat metric. We argue, as commonly accepted in such situations, that this parameter should be identified with the deformation of the \emph{complex} structure of the corresponding CY manifold. We confirm this conjecture below with detailed comments separately in the $N=1$, $N=2$, and $N>2$ cases.

This interpretation looks, however, rather surprising from the point of view of the original GLSM Lagrangian (\ref{wcpNN}), where the parameter in front of the twisted superpotential corresponds to the (complexified) parameter of the K\"ahler structure on the Higgs branch of the theory. Remember that in order to go to the Coulomb branch, we first had to set it to zero, since the Coulomb branch can only open up at $\beta=0$, and after integrating out matter multiplets, the inclusion of the superpotential already deforms a theory written in terms of different degrees of freedom. It is easy to check, for example, that the redefinition of $\beta$ by rescaling the $\Sigma$ field in the first case of the Higgs branch theory indeed changes the K\"ahler form (see e.g. (\ref{kin_super0}) or the original Lagrangian (\ref{wcpNN})), while in the effective theory with kinetic terms, given by (\ref{eff_kin_super}), a similar complex rescaling of $\Sigma$ does not affect the K\"ahler potential.

Remember also (see footnotes~\ref{foot:Coulomb}, \ref{foot:mass}) that, as we already discussed, the Coulomb branch is not well-defined in the 2d theory. Indeed, from the point of view of the \ntwo Liouville formulation, the Liouville potential does not allow the development of a nonvanishing value $\sigma\neq 0$. Classically, this is an exact statement, which actually states that there are no deformations of the complex structure, moving us away from the singular point. It turns out that this is also almost true quantum-mechanically (and we are going to discuss this in detail in particular cases) except for the $N=2$ conifold case, when the Liouville interaction corresponds to the operator from the spectrum of the theory, with a logarithmically normalizable wave-function. In this case, the 
$\tilde{\mu}$-deformation is allowed and corresponds to an existing deformation of the complex structure on the CY side.

\subsubsection{$N=1$ case}

As a first example of our equivalence, consider the simplest (naively?) case of the \wcpo model with just two complex
$n$ and $\rho$ fields. Its target space is one-dimensional complex or two-dimensional real, see \eqref{dimH}, and this
model, rewritten as a nonlinear sigma model (NLSM), was analyzed in \cite{AhRazSeib}. It has been shown numerically that the corresponding NLSM flows in the IR to a free theory of two real scalars plus fermion superpartners.

Let us check what one gets on the Liouville side. For $N=1$, there is no background charge in the \ntwo Liouville theory, since \eqref{Qexact} gives $Q=0$. To interpret this theory in the limit of vanishing radius $Q\to 0$ of the compact direction, it is easier to use its mirror description 
in terms of the \ntwo SL($2, \mathbb{R}$)/U(1) coset WZNW theory, where the relation \eqref{k_Q_relation} gives $k\to\infty$ for this case. Upon rescaling $\phi_c\to\phi_c/\sqrt{k}$, the cigar metric \eqref{BH} reduces at $k\to\infty$ to the flat two-dimensional target space
with constant dilaton \eqref{dilaton_cigar}. It shows that our equivalence relation \eqref{Qexact} perfectly works for the (opposite to the $N\to\infty$ limit) $N=1$ case.

\subsubsection{$N=2$: the conifold}

Let us now turn to the most important conifold case.
The conifold has two marginal deformations: of the K\"ahler form and complex structure, which cannot be switched on simultaneously. We have already discussed in 
Sect.~\ref{conifold}, that the parameter $\beta$ corresponds to the K\"ahler deformation of the resolved conifold, and now we argue (following \cite{GivKut,GivKutP}, where a similar problem was studied in the framework of AdS/CFT-like correspondence), that 
the Liouville parameter $\tilde{\mu}$ should be identified with the complex structure deformation on the CY side. This conclusion also follows from the analysis of the spectrum of the superconformal field theory, see e.g. \cite{ES}. Actually, as we see below, this is the only case when such deformation is essential from the target-space theory point of view.

A target-space argument, which supports such identification, looks as follows.
For $N=2$, the Liouville interaction $\sigma \sim e^{-\frac{\phi+iY}{Q}} = e^{-\frac{\phi+iY}{\sqrt{2}}}$ is a marginal primary vertex operator (\ref{vertex}) with $j=m=-1/2$, from the discrete spectrum (\ref{discrete}), associated with a massless physical state in 4d (since $\Delta = \frac12$ requires $p^2 = 0$ for the 4d momentum). Its wave function \eqref{psifun} is logarithmically normalizable, see \eqref{normalizable}. When raised to the exponent and included in the action (see \eqref{liouville}), the coefficient in front of this operator plays the role of the marginal deformation parameter of the conifold background. However, as already mentioned in Sect.~\ref{conifold} (see \cite{KSYconifold}), the K\"ahler form deformation modulus $\beta$ corresponds to the non-normalizable (quadratically divergent) deformation and should be associated with the coupling constant rather than with a dynamical state in the 4d theory, see \cite{GukVafaWitt} for the interpretation of non-normalizable states in CY compactifications. In contrast, the conifold complex structure modulus $\mu$ is logarithmically normalizable (on the border between normalizable and non-normalizable cases) and corresponds to a massless physical state in 4d \cite{KSYconifold}. We therefore identify
\beq
\tilde{\mu} \sim \mu  
\label{mu=tilde_mu}
\eeq
the parameter in front of the Liouville superpotential with the parameter of deformation of the complex structure.

To check the validity of the above identification, let us show that both sides in \eqref{mu=tilde_mu} transform in the same way with respect to the global symmetry group \eqref{global_sym}. In the $N=2$ case, the critical non-Abelian string has a massless state associated with the complex structure modulus $\mu$ of the conifold, see Sec.~\ref{conifold}. This state is a singlet with respect to both $SU(2)$ factors but has a baryonic charge $B=2$, see \eqref{n_rho_representations}. Eq.~\eqref{deformedconi} requires that $\mu$ transforms with respect to the $U(1)_B$ symmetry with the charge $B=2$.

On a Coulomb branch, this massless baryon is identified with the marginal primary operator \eqref{vertex} of the Liouville theory
with $j=-1/2$, $m=\pm 1/2$, $m \equiv m_L$. Moreover, the baryonic charge is related to shifts of the compact field $Y$, so that
$B=4m$ \cite{SYlittles,SYlittmult}, i.e. one has $B=\pm 2$ for the massless baryon with $m=\pm 1/2$.

To make the Liouville superpotential \eqref{liouville} invariant with respect to $U(1)_B$ symmetry, we require that the baryonic charge 
of $\tilde{\mu}$ should compensate the baryonic charge of the exponential with $m=-1/2$ in \eqref{liouville}. This gives  $B=2$ for the baryonic charge of $\tilde{\mu}$, i.e. the same value as the baryonic charge of $\mu$. 

Now let us turn to the 2d $R$-symmetry. Since the world-sheet \wcpN model is conformal, it has no chiral anomaly and therefore has two $R_{L,R}$-symmetries associated with rotations of $\theta^{+}$ and $\theta^{-}$. Normalizing the charge $R_L(\theta^{+}) = 1$, we see
that $Y$ should be shifted under the $R_L$ symmetry to make
the Liouville interaction \eqref{liouville} invariant. This gives the $R$-charge of the vertex operator \eqref{vertex} $R_L = -2m$ \cite{SYlittmult} (and similarly for the $R_R$
charge). Both $\mu$ and $\tilde{\mu}$ are neutral under $R$-symmetries. 

Finally, let us discuss the wave-function normalization.
To check the wave-function normalization, one has to relate the conifold radial coordinate
$r$ to the Liouville coordinate $\phi$. The Liouville superpotential
\eqref{WL} prevents the string from propagating to the region of large negative values, hence from \eqref{liouville} we can estimate that
\beq
\phi_{\text{min}} \sim \log{\tilde{\mu}}. 
\label{UVreg}
\eeq
On the other hand, from \eqref{deformedconi}, \eqref{tilder}, we see that 
$r_{\text{min}}^2 \sim |\mu|$. Upon identification \eqref{mu=tilde_mu}, this gives \cite{SYlittles}
\beq
\phi \sim \log{r^2}.
\label{phi_r}
\eeq
For $j=-1/2$, $m=\pm 1/2$, the wave function of the state \eqref{vertex} is $\phi$-independent, so that the norm is proportional to 
\beq
\phi_{\text{max}} - \phi_{\text{min}} \sim \log{\frac{r^2_{\text{max}}}{|\mu|}},
\label{norm}
\eeq
where $\phi_{\text{max}} \sim \log r_{\text{max}}^2$ is the IR regulator.
This is exactly what was found for the norm of the target-space scalar, associated with the modulus $\mu$ on the conifold \cite{KSYconifold}. 

To summarize, we start from the Higgs branch of the \wcpt model at $\beta \neq 0$, which geometrically corresponds to the resolved conifold. We move then to $\beta \to 0$, where the Coulomb branch of the \wcpt model opens up, and integrate over (massive at $\sigma \neq 0$) $n$ and $\rho$ fields. We then arrive at the Coulomb branch, described by the \ntwo Liouville theory. Here, the deformation operator, given by the same FI term, deforms the $\sigma$ space, rather than the $n$ and $\rho$ space, and geometrically corresponds to the deformed conifold with the complex structure parameter $\mu$. The whole process can be understood as a geometric transition, and the \ntwo Liouville theory gives a Lagrangian description of the theory on the deformed conifold.

Note that the deformation of the conifold complex structure, which becomes possible at $\beta = 0$, was not manifest in the original GLSM formulation of the \wcpN model, but now we see that it can be described in terms of the \ntwo Liouville theory. On the CY side, the parameter $\mu$ smooths the conifold singularity at small $r$, i.e. provides an ultraviolet regularization. In the Liouville theory, the Liouville superpotential at nonzero $\tilde{\mu}$ also provides a UV regularization, preventing the field $\phi$ from penetrating to the region of large negative values. With the identification \eqref{mu=tilde_mu}, the deformation of the conifold complex structure becomes manifest in the Liouville description.

\subsubsection{CY's with $N> 2$ }

In general, in the situation with $N\geq 3$, the Liouville operator does not coincide anymore with any of the primary vertex operators from the spectrum of Liouville theory. Indeed, for $\sigma = e^{-\frac{\phi+iY}{Q}}= e^{Q\left(-\frac{\phi}{Q^2}-\frac{iY}{Q^2}\right)}$, one can formally identify it with an element of the set \eqref{vertex} for
\beq
j=m=-\frac1{Q^2}= -\frac{k}{2}= -\frac1{2(N-1)}
\label{j_L}
\eeq
 which enters the spectrum \eqref{discrete} only for the $N=2$ case, and gives non-acceptable fractional values for $(j,m)$ if $N>2$. It would still be natural to identify the holomorphic Liouville superpotential with the deformation of the complex structure in the target-space theory, but the fact that the corresponding operator now drops out from the spectrum of the theory means that such analytic deformation no longer exists for $N\geq 3$. This conclusion can be supported by studying the correlation functions of the Liouville $\sigma$-fields, and already at the 2-point level, formula (\ref{refl_GK}) leads to a crucial difference between a constant (all \(\Gamma\)-functions cancel) for \(N=2\) and other cases.

 It is also easy to see that, for $N=3$, $Q=2$, $k=1/2$, the only primary operator \eqref{vertex} with conformal dimension $1/2$ has $j=-3/4$, $m=\pm 1/4$,
 \beq
 \Delta_{j=-\frac34,m=\pm \frac14 } =\frac12 
 \label{Delta-3/4}
 \eeq
 see \eqref{dimV}. However, this operator also does not belong to the discrete spectrum \eqref{discrete} and, in fact, has the form
 \eqref{genericj}, i.e. contains a rising exponent at $\phi\to\infty$ with $\tilde{j}=-1/4$
 and therefore is non-normalizable.

Below in Sec. \ref{moduli}, we study the \wcpN model for $N \geq 3$ and show that, in these cases, CY manifolds
are rigid and have no complex structure moduli. This means that the quantum Coulomb branch is not separated from the Higgs branch, and geometrically, we have a single target space.

\subsubsection{Black hole/string transition}
\label{phasetransition}

One can also argue the rigidity of the complex structure in the $N\geq 3$ case from the point of view of the black hole/string transition \cite{Susskind,HorowPolch}.
As we already mentioned, the mirror description 
of \ntwo Liouville theory is given by the supersymmetric version of Witten's two-dimensional black hole with
a semi-infinite cigar target space \cite{Wbh}, which is the SL(2,R)/U(1) coset WZNW theory 
\cite{GVafa,GivKut,MukVafa,OoguriVafa95}, see \eqref{BH}, \eqref{dilaton_cigar}.
The constant $\Phi_0$ in \eqref{dilaton_cigar} determines the mass of the black hole \cite{Wbh}.
In the Euclidean formulation, the compact dimension of the target space can be interpreted as a temperature circle with the temperature $(2\pi R)^{-1}$, where $R=\sqrt{2k}$ is the asymptotic radius of the cigar.

In string theory, at low temperatures, we have a well-defined black hole geometry with small $\alpha'\sim 1/k$ corrections. As the temperature grows above some critical value, the string's size exceeds its Schwarzschild radius and the black hole turns into an excited string \cite{HorowPolch}, 
similar behavior was found for the black hole \eqref{BH} with the linear dilaton \eqref{dilaton_cigar} in \cite{GivKutRabin}. It means that below
some critical value $k_c$, the $\alpha' \sim 1/k$ corrections grow, and the theory enters a strong coupling regime, where
the black hole, as a geometric object, no longer exists. 

In terms of the theory on the cigar \cite{GivKutRabin}, the Liouville superpotential
is a non-perturbative effect due to vortices \cite{HoriKapustin}, and at small $Q$ or large $k$, it represents a small correction.
As $k$ reduces, it becomes more and more important, and at $k_c=1$, it reaches the border between normalizable and non-normalizable operators, see \eqref{j_L}.
This suggests that the black hole \eqref{BH}, \eqref{dilaton_cigar} no longer exists for $k<1$ or $N>2$, and
we no longer have $\tilde{\mu}$ as a free deformation parameter in the \ntwo Liouville theory.

\section{A search for CY complex structure moduli}
\label{moduli}
\setcounter{equation}{0}

In this section, we show that our CY manifolds for $N\geq 3$ are rigid and do not have complex structure moduli. This can probably be extracted from the general discussion of rigid affine toric varieties in \cite{Alt} (see, for example, sect.~5 therein), but we prefer to give a direct explicit derivation.



\subsection{General setup}

Suppose some manifold \(\mathcal{M}\) of complex dimension \(d\) is defined by a set of polynomial equations
\begin{equation}
	\label{eq:1}
	F^i(\{w^{\alpha}\})=0,\quad i=1,\ldots, N_e, \quad \alpha=1,\ldots, N_v.
\end{equation}
Generally, this manifold should not be a complete intersection, i.e., \(N_v-N_e\leq d\).
Denote the ideal generated by \(F^i(\vec{w})\) by \(I\), so that functions on \(\mathcal{M}\) are
\begin{equation}
	\label{eq:7}
	Fun(\mathcal{M})=\mathbb{C}[\{w^{\alpha}\}]/I.
\end{equation}

\subsection{Deformations}

To consider the infinitesimal deformations of this manifold, let us add some r.h.s. to the equations \eqref{eq:1}
\begin{equation}
	\label{eq:2}
	F^i(\vec{w})=\epsilon \delta F^i(\vec{w})+O(\epsilon^2).
\end{equation}
The first possible problem is that the dimension of the manifold can drop, $\dim \mathcal{M}_\epsilon < \dim \mathcal{M}$.
To check the dimension, consider \eqref{eq:2} in the vicinity of a point $\vec{w}$ of the deformed manifold:
\begin{equation}
	\label{eq:3}
	F^i(\vec{w}+\epsilon \delta \vec{w})=\epsilon \delta F^i(\vec{w}+\epsilon \delta \vec{w})+O(\epsilon^2),
\end{equation}
which gives, to leading order,
\begin{equation}
	\label{eq:eqPerturbative}
	F^i(\vec{w})=0,\qquad \sum_{\alpha}\frac{\partial F^i(\vec{w})}{\partial w^\alpha}\delta w^\alpha=\delta F^i(\vec{w}).
\end{equation}
We know that at the generic point, \(\dim\ker \frac{\partial F^i(\vec{w})}{\partial w^\alpha}=d\), since it corresponds to the tangent space to \(\mathcal{M}\).
Hence, the second equation in \eqref{eq:eqPerturbative}, which is linear in \(\delta w^{\alpha}\), either has no solutions or has a \(d\)-dimensional space of solutions (corresponding to the tangent space to the deformed manifold, $\dim \mathcal{M}_\epsilon$).
The second option is only possible if the r.h.s. is in the image of the operator in the l.h.s.:
\begin{equation}
	\label{eq:image}
	\forall \vec{w}\in \mathcal{M}: \delta F^i(\vec{w})\in \im \frac{\partial F^i(\vec{w})}{\partial w^\alpha},
\end{equation}
which is a non-trivial condition for deformations \(\delta F^i(\vec{w})\).
It is therefore convenient to define an operator \(D_0(\vec{\rho},\vec{n})\) 
\begin{equation}
	\label{eq:9}
	D_0(\vec{\rho},\vec{n})=\left.\frac{\partial F^i(\vec{w})}{\partial w^\alpha}\right|_{w^{ij}=\rho^in^j}
\end{equation}
which lives only on \(\mathcal{M}\)~\footnote{In this way, one immediately gets rid of the trivial deformations of the form
	\begin{equation}
		\label{eq:13}
		\delta F^i(\vec{w}) = \epsilon \sum_j c^i_j(\vec{w}) F_j(\vec{w}).
	\end{equation}
}.

Condition \eqref{eq:image} is not constructive, and it is better
to have some description of the space of deformations \(\delta F^i(\vec{w})\) (the image of \(D_0\)) as the kernel of some operator, namely
\begin{equation}
	\label{eq:imageDescription1}
	\vec{v}\in \im D_0(\vec{\rho},\vec{n}) \Leftrightarrow D_1(\vec{\rho},\vec{n})\vec{v}=0.
\end{equation}
The rank of the matrix \(\rk D_0(\vec{\rho},\vec{n})=N_v-d\) is constant everywhere except at \(\vec{w}=0\), since the only singularity is at the origin, and
in order to get a precise description \eqref{eq:imageDescription1}, one should guarantee that \(\dim\ker D_1(\vec{\rho},\vec{n})=N_v-d\) everywhere except at zero.

Notice that the matrix \(D_1\) can be chosen in many different ways.
It turns out that \(D_1\) can be constructed to be linear in \(\vec{n}\) and \(\vec{\rho}\).


Non-trivial deformations of the manifold \(\mathcal{M}\) can be found as the factorization of all deformations or  \(\ker D_1\) by trivial ones,
(which can be removed by an appropriate change of coordinates \(\delta w^\alpha=\epsilon c^{\alpha}(\vec{w})\))
\begin{equation}
	\label{eq:12}
	\delta F^i(\vec{w})=\epsilon\sum_{\alpha} D_0(\vec{w})^i_{\alpha}c^{\alpha}(\vec{w}).
\end{equation}
and lie in \(\im D_0\).

Let us introduce the spaces
\begin{equation}
	\label{eq:15}
	V_{k,l}=\{f\in \mathbb{C}[\vec{\rho},\vec{n}]| \deg_{\vec{\rho}}f=k, \deg_{\vec{n}}f=l\}.
\end{equation}
Since \(\delta F^i(\vec{w})\in \mathbb{C}[\vec{w}]\), after specialization, it becomes an element of \(\oplus_{k=0}^{\infty} V_{k,k}\).
Notice also that \(D_0\) preserves the degree, so it is natural to define its restrictions
\begin{equation}
	\label{eq:14}
	D_0[k]=\left.D_0(\vec{\rho},\vec{n})\right|_{V_{k,k}}, \ \ \ \ \  D_1[k]=\left.D_1(\vec{\rho},\vec{n})\right|_{V_{k,k}}.
\end{equation}
These operators act as:
\be
	\label{eq:23}
	D_0[k]: V_{k,k}\otimes \mathbb{C}^{N_v}\to V_{k+1,k+1}\otimes \mathbb{C}^{N_e},
	\\
	D_1[k]: V_{k,k}\otimes \mathbb{C}^{N_e}\to \left(V_{k+1,k}\oplus V_{k,k+1}\right)\otimes \mathbb{C}^{N_e}.
\ee
and one can describe the deformation space now as
\begin{equation}
	\label{eq:17}
	H[k]=\ker D_1[k]/\im D_0[k-1].
\end{equation}
The dimension of this space can be computed as
\be
	\label{eq:18}
	\dim H[k]=\dim\ker D_1[k]-\dim\im D_0[k-1] = 
	\\
	= \dim V_{k,k}-\rk D_1[k]-\rk D_0[k-1],
\ee
since \(\dim\im D_0[k]=\rk D_0[k]\) and \(\dim\ker D_1[k]=\dim V_{k,k}-\rk D_1[k]\).


\subsection{General \(N\geq 3\) proof}
\label{ss:deform}

In our case of interest \(N_v=N^2\) coordinates
\begin{equation}
	\label{eq:4}
	w^{ij}=\rho^in^j,\quad i,j=1,\ldots, N.
\end{equation}
are constrained by \(N_e=\left(\frac{N(N-1)}{2}\right)^2\) equations~\footnote{These equations have obvious symmetries:
	\begin{equation}
		\label{eq:Fsymmetry}
		F^{[ij][kl]}=-F^{[ji][kl]}=-F^{[ij][lk]}=F^{[ji][lk]}.
\end{equation}
}
\begin{equation}
	\label{eq:6}
	F^{[ij][kl]}=w^{ik}w^{jl}-w^{jk}w^{il} = 0.
\end{equation}
The tangent space at a generic point can be defined by \(D_1(\vec{\rho},\vec{n})\vec{v}=0\) for the operator:
\be
	D_1(\vec{n},\vec{\rho})_{[ij][kl]}^{[i_1,i_2,i_3]\rho}=\sum_{\sigma\in S_3}(-1)^\sigma \rho^{\sigma(i_1)} \delta_{i\sigma(i_2)}\delta_{j\sigma(i_3)}
\\
	D_1(\vec{n},\vec{\rho})_{[ij][kl]}^{[i_1,i_2,i_3]n}=\sum_{\sigma\in S_3}(-1)^\sigma n^{\sigma(i_1)} \delta_{k\sigma(i_2)}\delta_{l\sigma(i_3)}
\label{eq:imN1}
\ee
It is easy to check that \(D_1(\vec{\rho},\vec{n})D_0(\vec{\rho},\vec{n})=0\), e.g. for the first equation
\begin{equation}
	\label{eq:37}
	\sum_{i_1i_2i_3} \epsilon^{i_1i_2i_3}\rho^{i_1}\frac{\partial (w^{i_2j}w^{i_3k}-w^{i_3j}w^{i_2k})}{\partial w^{ak}}=\sum_{i_1i_2i_3} (\epsilon^{i_1i_2a}\rho^{i_1}w^{i_2j}- \epsilon^{i_1ai_3}\rho^{i_1}w^{i_3j}),
\end{equation}
so that both terms on the r.h.s. vanish after the substitution \(w^{ij}=\rho^in^j\) due to the antisymmetry.

Equations \(D_1(\vec{\rho},\vec{n})\vec{v}=0\) for the operator \eqref{eq:imN1} have the form
\begin{equation}\begin{gathered}
		\label{eq:imD1}
		\rho^{i_1}v^{[i_2i_3][j_1j_2]}+\rho^{i_2}v^{[i_3i_1][j_1j_2]}+\rho^{i_3}v^{[i_1i_2][j_1j_2]}=0,\\
		n^{j_1}v^{[i_1i_2][j_2j_3]}+n^{j_2}v^{[i_1i_2][j_3j_1]}+n^{j_3}v^{[i_1i_2][j_1j_2]}=0.
\end{gathered}\end{equation}
and should hold for all \(i_1<i_2<i_3\) and \(j_1<j_2<j_3\).

To solve \eqref{eq:imD1} in terms of polynomial functions we use the following
\begin{lemma}
	\label{thm:lemma1}
	Let \(M\) polynomial functions \(f_1,\ldots, f_M \in R[x_1,\ldots x_M]\), where \(R\) is an arbitrary ring (for example, the ring of polynomials in some other variables), satisfy 
	\begin{equation}
		\label{eq:orthogonalityEq}
		\sum_{k=1}^Mx_kf_k=0.
	\end{equation}
	Then \(\exists \Omega_{km}\in R[x_1,\ldots ,x_M]\), restricted by the antisymmetry 
	\begin{equation}
		\label{eq:38}
		\Omega_{km}=-\Omega_{mk},
	\end{equation}
	such that
	\begin{equation}
		\label{eq:orthogonalitySol}
		f_k=\sum_{m=1}^M\Omega_{km}x_m.
	\end{equation}
\end{lemma}
In other words, \eqref{eq:orthogonalitySol} gives general solution to \eqref{eq:orthogonalityEq}.

\textit{Proof:} We prove it by induction in \(M\); for \(M=1\) the statement is trivial.
Suppose that it is true for \(M-1\).
Rewrite \eqref{eq:orthogonalityEq} as
\begin{equation}
	\label{eq:31}
	x_Mf_M=-\sum_{k=1}^{M-1}x_kf_k.
\end{equation}
so that it becomes clear, that \(f_M\) cannot contain monomials \(x_M^n\), since they are absent in the rhs. Therefore
\begin{equation}
	\label{eq:solfM}
	f_M=\sum_{k=1}^{M-1}\omega_kx_k
\end{equation}
for some \(\omega_k\in R[x_1,\ldots ,x_M]\).
Substituting it into \eqref{eq:orthogonalityEq}:
\begin{equation}
	\label{eq:36}
	\sum_{k=1}^{M-1}x_k(f_k+\omega_kx_M)=0
\end{equation}
and using the induction assumption, one can solve this equation as:
\begin{equation}
	\label{eq:solfk}
	f_k+\omega_kx_M=\sum_{k=1}^{M-1}\Omega_{km}x_m.
\end{equation}
Formulas \eqref{eq:solfM} and \eqref{eq:solfk} together give \eqref{eq:orthogonalitySol}, if one defines
\begin{equation}
	\label{eq:33}
	\Omega_{Mk}=\omega_k,\qquad \Omega_{kM}=-\omega_k,\qquad \Omega_{MM}=0,
\end{equation}
which completes the proof. \hfill \(\square\)

\vspace{0.5cm}
Using Lemma~\ref{thm:lemma1} one can solve the first equation \eqref{eq:imD1} (with fixed \(i_1, i_2, i_3\)):
\begin{equation}\begin{gathered}
		\label{eq:8}
		v^{[i_2i_3][j_1j_2]}=c \rho^{i_2}-b \rho^{i_3},\quad v^{[i_3i_1][j_1j_2]}=a \rho^{i_3}-c \rho^{i_1},\quad v^{[i_1i_2][j_1j_2]}=b \rho^{i_1}-a \rho^{i_2}.
	\end{gathered}
\end{equation}
Using antisymmetry in \(i_k\) and the fact that it holds for arbitrary \(i_1, i_2, i_3\) we conclude that
\begin{equation}
	\label{eq:35}
	v^{[i_1i_2][j_1j_2]}=\rho^{i_1}a^{i_2[j_1j_2]}-\rho^{i_2}a^{i_1[j_1j_2]}.
\end{equation}
Then we solve the second equation in~\eqref{eq:imD1} in the same way, which turns it into an equation for \(a\):
\begin{equation}
	\label{eq:39}
	a^{i_1[j_1j_2]}=d^{i_1j_1}n^{j_2}-d^{i_1j_2}n^{j_1},
\end{equation}
therefore
\begin{multline}
	\label{eq:40}
	v^{[i_1i_2][j_1j_2]}=d^{i_1j_1}\rho^{i_2}n^{j_2}-d^{i_2j_1}\rho^{i_1}n^{j_2}+d^{i_2j_2}\rho^{i_1}n^{j_1}-d^{i_1j_2}\rho^{i_2}n^{j_1}=
	\\
	=
	d^{i_1j_1}w^{i_2j_2}-d^{i_2j_1}w^{i_1j_2}+d^{i_2j_2}w^{i_1j_1}-d^{i_1j_2}w^{i_2j_1}=
	\\
	=
	\left( d^{i_1j_1}\frac{\partial }{\partial w^{i_1j_1}} + d^{i_1j_1}\frac{\partial }{\partial w^{i_1j_1}}+d^{i_2j_2}\frac{\partial }{\partial w^{i_2j_2}}+d^{i_1j_2}\frac{\partial }{\partial w^{i_1j_2}} \right)
	\\
	(w^{i_1j_1}w^{i_2j_2}-w^{i_2j_1}w^{i_1j_2}).
\end{multline}
The last expression, adding trivially vanishing terms, can be rewritten as
\begin{equation}
	\label{eq:41}
	v^{[i_1i_2][j_1j_2]}=\sum_{ij}d^{ij}\frac{\partial }{\partial w^{ij}}(w^{i_1j_1}w^{i_2j_2}-w^{i_2j_1}w^{i_1j_2})=
	\sum_{ij}d^{ij}\frac{\partial F^{[i_1i_2][j_1j_2]}}{\partial w^{ij}},
\end{equation}
or just 
\begin{equation}
	\label{eq:42}
	\vec{v}=D_0\vec{d},
\end{equation}
so necessarily \(\vec{v}\in \im D_0\), and \(\ker D_1=\im D_0\). We conclude therefore, that all manifolds for $N>3$ do not have non-trivial deformations.
For the illustration purposes we collected some explicit formulas for the \(N=3\) case in Appendix~\ref{ap:N3}.

\subsection{Conifold}

In the exceptional \(N=2\) conifold case, where the general proof does not work, one has a single equation:
\begin{equation}
	\label{eq:19}
	F=w^{11}w^{22}-w^{12}w^{21}=0,
\end{equation} 
which is obviously solved by \(w^{ij}=\rho^i n^j\). The Jacobian \eqref{eq:9} (with the order of variables \(w^{11}\), \(w^{12}\), \(w^{21}\), \(w^{22}\)) is:
%
\begin{equation*}
\begin{aligned}
	D_0(\vec{\rho},\vec{n})
		&= \begin{pmatrix}
				w^{22} \,, & -w^{21} \,, & -w^{12} \,, & w^{11} 
			\end{pmatrix} \Big|_{w^{ij}=\rho^in^j}
			\\
		&= \begin{pmatrix}
				\rho^2n^2 \,, & -\rho^2n^1 \,, & -\rho^1n^2 \,, & \rho^1n^1
			\end{pmatrix}
\end{aligned}
\end{equation*}
The operator (\ref{eq:imN1}) vanishes identically here, i.e., its kernel contains all polynomial functions \(\ker D_1=\mathbb{C}[\vec{n},\vec{\rho}]^{\deg_n=\deg_{\rho}}\). Thus, 
the kernel \(\ker D_1[k]\) is spanned by vectors \(
\begin{pmatrix}
	x & x
\end{pmatrix}\), with \(x\in V_{k,k}\), or being an arbitrary linear combination of \((n^1)^a(n^2)^{k-a}(\rho^1)^b(\rho^2)^{k-b}\). However, in this particular case, \(\im D_0=\mathbb{C}[\vec{n},\vec{\rho}]^{\deg_n=\deg_{\rho}\ge 1}\) is different, since
the image \(\im D_0[k-1]\) is spanned by vectors \(
\begin{pmatrix}
	y & y
\end{pmatrix}\), where \(y=t_{11}\rho^2n^2-t_{12}\rho^2n^1-t_{21}\rho^1n^2+t_{22}\rho^1n^1\) with all \(t_{ij}\in V_{k-1,k-1}\).
We see that all \(V_{k,k}\) are reproduced by choosing appropriate \(t_{ij}\), except for the subspace with \(k=0\).

This means that \(H[k]=\ker D_1[k]/\im D_0[k-1]=0\) for \(k\geq 1\), and \(H[0]=\ker D_1[0]=\mathbb{C}\). 
This corresponds to the complex parameter $\mu$ in \eqref{deformedconi}.


\section{Conclusions}
\label{conclusions}
\setcounter{equation}{0}

In this paper, we have demonstrated that the Coulomb branch of \wcpN models can be effectively described by \ntwo Liouville theory.
We found the exact formula for the Liouville background charge
$Q=\sqrt{2(N-1)}$, requiring that the central charges of both theories coincide, and we have demonstrated that large-$N$ calculation confirms 
this formula in the leading order.

We have also identified the coefficient $\tilde{\mu}$ in front of the Liouville superpotential with the parameter of deformation of the complex structure
of the corresponding CY manifold. However, except for the $N=2$ conifold case, this space is empty (in contrast to the case of K\"ahler deformations of Higgs branches, existing for arbitrary $N$), which is related to the fact that the Liouville superpotential does not correspond to any normalizable state in the spectrum of the theory.
Qualitatively, this means that Coulomb and Higgs branches are not separated in these models, and the ground state wave function is spread over the whole target space.

The conifold case ($N=2$) is special. The Higgs and the Coulomb branches of the \wcpt model are geometrically distinct and correspond to the resolved and deformed conifold, respectively. In this case, the identification of $\tilde{\mu}$ with the conifold complex structure modulus $\mu$ confirms the proposal of \cite{GivKut,GivKutP}.
For $N>2$ cases, the absence of marginal primaries on the Liouville side matches 
with our results, which show the absence of complex structure moduli for all $N\geq 3$ cases.

\section*{Acknowledgments}

The authors are grateful to M.~Bershtein, G.~Korchemsky, A.~Litvinov, N.~Nekrasov and M.~Shifman for valuable discussions. 
The work of E.I.,  I.M. and A.Y. was supported by the Foundation for the Advancement of Theoretical Physics and Mathematics ''BASIS'',  Grant No. 22-1-1-16.
The work of E.I. was also partly supported by the FY2021-SGP-1-STMM Grant No. 021220FD3951 at Nazarbayev University.

%
%

\appendix

\section{Derivation of the one loop effective action}
\label{sec:loops}

\begin{figure}[h]
    \centering
    \begin{subfigure}[t]{0.45\textwidth}
        \centering
        \includegraphics[width=\textwidth]{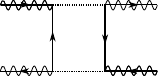}
    \end{subfigure}%
    ~ 
    \begin{subfigure}[t]{0.24\textwidth}
        \centering
        \raisebox{7pt}{\includegraphics[width=\textwidth]{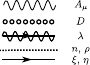}}
    \end{subfigure}%
\caption{
	The four-gaugino diagram and notation. 
	The diagram can be computed at zero external momentum.
	Note that in the chiral vertices the fermionic currents (big arrows) face to each other.
	}
\label{fig:diag_Lam4}
\end{figure}

\begin{figure}[h]
    \centering
    \begin{subfigure}[t]{0.5\textwidth}
        \centering
        \includegraphics[width=\textwidth]{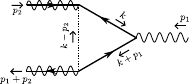}
    \end{subfigure}%
    ~ 
    \begin{subfigure}[t]{0.5\textwidth}
        \centering
        \includegraphics[width=\textwidth]{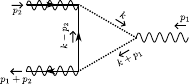}
    \end{subfigure}%
    
    \begin{subfigure}[t]{0.45\textwidth}
        \centering
        \includegraphics[width=\textwidth]{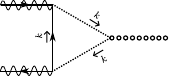}
    \end{subfigure}%
\caption{Diagrams for the Yukawa coupling of the gaugino to gauge bosons. 
For the notation, see Fig.~\ref{fig:diag_Lam4}. 
The two diagrams on the top are computed to the first power in the photon momentum \(p_1\) (they differ by internal lines in the triangles). 
The last diagram can be computed at zero external momentum.
Note that in the chiral vertices, the fermionic currents (big arrows) either face or oppose each other.
Little arrows denote momentum flow.}
\label{fig:diag_LamLamA}
\end{figure}

Here we give a brief overview of how to derive the effective action \eqref{eff_kin}.
We start by restoring the fermionic part of the $\mathbb{WCP}(N,N)$ model action (in Minkowski spacetime; for the bosonic part, see \eqref{wcpNN}):
\begin{equation}
	\begin{aligned}
		S_\text{ferm} 
		&= \int d^2 x \Bigg\{ 
		\frac{1}{e_{0}^{2}} i \bar{\lambda}_{R} (\partial_{0} +\partial_{1}) \lambda_{R} 
		+ \frac{1}{e_{0}^{2}} i \bar{\lambda}_{L} (\partial_{0} -\partial_{1}) \lambda_{L} \\
		&+i \bar{\xi}_{R} (\partial_{0} +\partial_{1}) \xi_{R} + i \bar{\xi}_{L} (\partial_{0} -\partial_{1}) \xi_{L} \\
		&+ i \bar{\eta}_{R} (\partial_{0} +\partial_{1}) \eta_{R} + i \bar{\eta}_{L} (\partial_{0} -\partial_{1}) \eta_{L}\\
		&- \sqrt{2}\sigma \bar{\xi}_{R i} \xi_{L}^{i} 
		- \sqrt{2} \bar{\sigma} \bar{\xi}_{L i} \xi_{R}^{i} 
		+ \sqrt{2} \sigma \bar{\eta}_{R i} \eta_{L}^{i}
		+\sqrt{2}\bar{\sigma} \bar{\eta}_{L i} \eta_{R}^{i} \\
		&+i \sqrt{2} \bar{n}_{i}\left( \xi_{R}^{i}\lambda_{L}- \xi_{L}^{i}\lambda_{R}\right)
		+i \sqrt{2} n^{i}\left(\bar{\lambda}_{R} \bar{\xi}_{L i}-\bar{\lambda}_{L} \bar{\xi}_{R i}\right)\\
		&-i \sqrt{2} \bar{\rho}_{i}\left( \eta_{R}^{i}\lambda_{L}- \eta_{L}^{i}\lambda_{R}\right)
		-i \sqrt{2} \rho^{i}\left(\bar{\lambda}_{R} \bar{\eta}_{L i}-\bar{\lambda}_{L} \bar{\eta}_{R i}\right)
		\Bigg\}\\
=& \int d^2 x \Bigg\{ 
		\frac{1}{e_{0}^{2}} i\bar{\Lambda}\gamma^{\mu}\partial_{\mu}\Lambda   
		+ i\bar{\Xi}\gamma^{\mu}\nabla_{\mu}\Xi +i\bar{H}\gamma^{\mu}\nabla_{\mu}H \\
		&- \bar{\Xi} M(\sigma) \Xi
		+ \sqrt{2} \bar{n}_{i}{\bar{\Xi}}^{*}\Lambda
		- \sqrt{2} n^{i}\bar{\Lambda}\Xi^{*}\\
		&+ \bar{H} M(\sigma) H
		- \sqrt{2} \bar{\rho}_{i}{\bar{H}}^{*}\Lambda
		+ \sqrt{2} \rho^{i}\bar{\Lambda}H^{*} 
		\Bigg\} 
		\,,
	\end{aligned}
	\label{wcpNN_ferm}
\end{equation}
where $\xi_{R,L}^{i}$, $\eta_{R,L}^{i}$ are the superpartners of $n_{i}$, $\rho_i$ respectively, and 
we use the following representation:
\begin{equation}
	\begin{aligned}
		\gamma^{0}=
		&\left(\begin{array}{ll}
			0 & -i \\
			i & 0
		\end{array}\right), \quad 
		\gamma^{1}=\left(\begin{array}{cc}
			0 & i \\
			i & 0
		\end{array}\right), \quad
		\gamma^{\text {chir }}=\gamma^{0} \gamma^{1}=\left(\begin{array}{cc}
			1 & 0 \\
			0 & -1
		\end{array}\right) \,, \\
		&\Xi=\left(\begin{array}{c}
			\xi_{R} \\
			\xi_{L}
		\end{array}\right) \,, \quad 
		\Lambda=\left(\begin{array}{c}
			\lambda_{R} \\
			\lambda_{L}
		\end{array}\right) \,, \quad
		H=\left(\begin{array}{c}
			\eta_{R} \\
			\eta_{L}
		\end{array}\right) \,.
		\label{two_component_spinors}
	\end{aligned}
\end{equation}
with the Dirac conjugation defined as $\bar{\Xi}=\Xi^{\dagger} \gamma_{0}=\left(i\bar{\xi}_{L}, -i\bar{\xi}_{R}\right)$, and the same for $\Lambda$ and $H$.
The complex conjugated spinors (i.e. without transposing and $\gamma_0$) are denoted as
\begin{equation}
	\Xi^* = \left(\begin{array}{c}
		\bar{\xi}_{R} \\
		\bar{\xi}_{L}
	\end{array}\right) \,, \quad 
	\bar{\Xi}^* = \left(- i \xi_{L}, i \xi_{R}\right) \,,
\end{equation}
and the same for $H^*$, $\bar{H}^*$, and
the fermion mass matrix is given by
\begin{equation}
	M(\sigma) = 
	\begin{pmatrix}
		- i \sqrt{2} \bar{\sigma} & 0 \\
		0 & i \sqrt{2} \sigma 
	\end{pmatrix}
	=
	- i \sqrt{2} \left( \gamma^{\text{chir}} \, \Re\sigma  - i \, \mathbb{I} \, \Im\sigma  \right)
	\label{MMat}
\end{equation}
where $\mathbb{I}$ is an identity matrix.
The Feynman rules are easily read off the action \eqref{wcpNN_ferm}. Note that it contains chiral vertices such as $\bar{n}_{i}{\bar{\Xi}}^{*}\Lambda$, these vertices show up in Feynman graphs together with adjacent fermionic arrows facing each other (or turning away from each other).

The effective action \eqref{eff_kin} comes from one-loop diagrams. 
The kinetic terms for the gauge hypermultiplet fields $A_\mu$, $\lambda_{L,R}$, $\sigma$, $D$ come from the diagrams as on Fig.~\ref{fig:esigma}, with  $n_i$, $\rho_i$, $\xi_i$, $\eta_i$ propagating in the internal loop. 
T	hese  diagrams were computed in e.g. \cite{W79,SYhetN,Ievlev:2020qcy} (summation over all flavors yields a coefficient $N_F=2N$).
The Yukawa terms in \eqref{eff_kin} come from the triangle diagrams shown on Fig.~\ref{fig:diag_LamLamA}, see Fig.~\ref{fig:diag_Lam4} for the notation for propagator lines. We calculate these diagrams now to the lowest order in external momenta.

At first glance, it seems that the total contribution of triangle diagrams from Fig.~\ref{fig:diag_LamLamA} vanishes after suming over $N$ flavors with charges $\mathtt{Q}=+1$ and $N$ flavors with charges $\mathtt{Q}=-1$, since they are  proportional to $\mathtt{Q}^3$. 
However, it follows from the action \eqref{wcpNN_ferm} that the fermionic propagator depends non-trivially on the charge:
%
\begin{equation}
	\begin{aligned}
		\langle \Xi \bar{\Xi} \rangle, \ \langle H \bar{H} \rangle
		&= \frac{i}{ \slashed{p} - \mathtt{Q} \cdot M(\sigma)  } 
		= i \frac{\slashed{p} + \mathtt{Q} \cdot M(\sigma)^\dagger}{ p^2 -  2|\sigma|^2   } \,,
	\\
		\langle \bar{\Xi}^* \Xi^* \rangle, \ \langle \bar{H}^* H^* \rangle
		&= \frac{i}{ - \slashed{p} - \mathtt{Q} \cdot M^\dagger(\sigma)  } 
		= i \frac{ - \slashed{p} + \mathtt{Q} \cdot M(\sigma)}{ p^2 -  2|\sigma|^2   } \,,
	\end{aligned}
\end{equation}
Hence, the triangle diagrams on Fig.~\ref{fig:diag_LamLamA} apart from the terms $\sim \mathtt{Q}^3$ also contain the terms $\sim \mathtt{Q}^4$ which do not cancel out.
The diagrams with an external photon leg are calculated to the first order in the photon momentum $p_1$ and give contributions $\sim \epsilon_{\mu \nu} p_1^\mu A^\nu$. 
At this order, dependence on $p_2$ cancels when we take the sum of all diagrams, i.e. we get the terms
\begin{equation}
	- \frac{N_{F}}{2} \frac{ iF_{01} }{4 \pi |\sigma|^2}
	\left( \frac{ {\lambda}_{R} \bar{\lambda}_{L} }{\sqrt{2} \sigma } - \frac{ {\lambda}_{L} \bar{\lambda}_{R} }{\sqrt{2} \bar{\sigma} }   \right) \,
	\label{effact_1}
\end{equation}
in the effective action, where again \(N_F = 2 N\).

The diagram with a \(D\)-field at the external leg is calculated at zero external momentum and yields
\begin{equation}
	- \frac{N_{F}}{2}\frac{D}{4 \pi |\sigma|^2} 	
	\left(\frac{ {\lambda}_{R} \bar{\lambda}_{L} }{\sqrt{2} \sigma } + \frac{ {\lambda}_{L} \bar{\lambda}_{R} }{\sqrt{2} \bar{\sigma} }\right)
	\label{effact_2}
\end{equation}

The four-legged diagram on Fig.~\ref{fig:diag_Lam4} is a source of the four-fermion terms. 
Performing the calculation in terms of the two-component spinors \eqref{two_component_spinors}, one has to take into account a symmetry factor \(1/2\), related to the fact that this diagram has two independent fermion current structures and effectively yields a square of a bifermion current. 
With this symmetry factor, the diagram equals:
\begin{equation}
	- \frac{iN_{F}{\gamma}^{\mu}_{ \dot \alpha  \alpha}{\gamma}_{\mu \dot \beta  \beta}}{24 \pi |M(\sigma)|^4}
	+ \frac{ iN_{F}{ M (\sigma) }_{ \dot \alpha  \alpha}{ M (\sigma) }_{ \dot \beta  \beta} }{6\pi |M(\sigma)|^6} \,.
\end{equation}
%
%
%
Contractions with \(\Lambda\)'s, using \eqref{MMat} and
\begin{equation}
	\begin{aligned}
		{\bar \Lambda}_{\dot \alpha}\gamma_{\text {chir }}^{\dot \alpha \alpha}	{\Lambda}_{ \alpha}
		&= - i( {\lambda}_{R} \bar{\lambda}_{L} + {\lambda}_{L} \bar{\lambda}_{R} ) \,, \\
		{\bar \Lambda}_{\dot \alpha} \mathbb{I}^{\dot \alpha \alpha}	{\Lambda}_{ \alpha} 
		&= - i( {\lambda}_{R} \bar{\lambda}_{L} - {\lambda}_{L} \bar{\lambda}_{R} ) \,, \\
		{\bar \Lambda}_{\dot \alpha} {\gamma}^{\mu \dot \alpha  \alpha}{\Lambda}_{ \alpha} 	{\bar \Lambda}_{\dot \beta} {\gamma}_{\mu}^{ \dot \beta  \beta} {\Lambda}_{ \beta} 
		&= -4 {\lambda}_{R} \bar{\lambda}_{L} {\lambda}_{L}  \bar{\lambda}_{R} \,,
	\end{aligned}
\end{equation}
give the four-fermion term
\begin{equation}
	\frac{N_{F}}{2}\frac{ {\lambda}_{L} \bar{\lambda}_{R} {\lambda}_{R} \bar{\lambda}_{L} }{4 \pi |\sigma|^4} \,.
	\label{effact_3}
\end{equation}
Collecting all pieces \eqref{effact_1}, \eqref{effact_2}, and \eqref{effact_3} together, one gets all interaction terms in the effective action \eqref{eff_kin}.

\section{$N=3$ example}
\label{ap:N3}

This is the simplest illustration of the general construction from Sec.~\ref{ss:deform},
when a submanifold \(\mathcal{M}\subset\mathbb{C}^{N_v}=\mathbb{C}^9\) given by
\begin{equation}
	w^{ij}=\rho^in^j,\quad i,j=1, 2, 3,
\end{equation}
of dimension \(d=5\) is defined by a system of equations
\begin{equation}
	\label{eq:Fequations}
	F^{kk'}(\vec{w})=w^{ii'}w^{jj'}-w^{ji'}w^{ij'}=0,
\end{equation}
where \(i, j, k\) and \(i', j', k'\) are triples of all different numbers. Here
we have \(N_v=N_e=9\) and \(d=5\), so that \(\rk D_0 =N_v-d=4\).

It is easy to understand that the equations \eqref{eq:Fequations} actually describe \(\mathcal{M}\) without any extra components.

First, in the vicinity of a point \(\vec{n} = \begin{pmatrix}1 & 0 & 0\end{pmatrix}\), \(\vec{\rho} = \begin{pmatrix}1 & 0 & 0\end{pmatrix}\), the linearized expansion of \eqref{eq:Fequations} around this point yields a system of equations \(\delta w^{23} = \delta w^{32} = \delta w^{22} = \delta w^{33} = 0\), which defines the tangent space of the correct dimension \(d = 5\).

Second, the group \(GL(3) \times GL(3) / GL(1)^{\diag}\) acts transitively on \(\mathcal{M}\) except at zero, and linearly on the equations \eqref{eq:Fequations}. This means that all points of \(\mathcal{M}\), except zero, are equivalent, so the tangent space defined by \eqref{eq:Fequations} has dimension \(d = 5\) at all points of \(\mathcal{M}\), except zero.

Introducing the following ordering of the sets of equations \(F^{kk'}\) and variables \(w^{ii'}\)
\be
\label{eq:24}
\begin{pmatrix}
	F^i
\end{pmatrix}
=
\begin{pmatrix}
	F^{11} & F^{12} & F^{13} & F^{21} & F^{22} & F^{23} & F^{31}& F^{32} & F^{33}
\end{pmatrix},
\\
\begin{pmatrix}
	w_\alpha
\end{pmatrix}
=
\begin{pmatrix}
	w^{11} & w^{12} & w^{13} & w^{21} & w^{22} & w^{23} & w^{31}& w^{32} & w^{33}
\end{pmatrix},
\ee
the matrix (\ref{eq:9}) acquires the form
\begin{equation}
	\label{eq:5}
	D_0=
	\begin{pmatrix}
		0& 0& 0& 0& w^{33}& -w^{32}& 0& -w^{23}& w^{22}\\
		0& 0& 0& w^{33}& 0& -w^{31}& -w^{23}& 0& w^{21}\\
		0& 0& 0& w^{32}& -w^{31}& 0& -w^{22}& w^{21}& 0\\
		0& w^{33}& -w^{32}& 0& 0& 0& 0& -w^{13}& w^{12}\\
		w^{33}& 0& -w^{31}& 0& 0& 0& -w^{13}& 0& w^{11}\\
		w^{32}& -w^{31}& 0& 0& 0& 0& -w^{12}& w^{11}& 0\\
		0& w^{23}& -w^{22}& 0& -w^{13}& w^{12}& 0& 0& 0\\
		w^{23}& 0& -w^{21}& -w^{13}& 0& w^{11}& 0& 0& 0\\
		w^{22}& -w^{21}& 0& -w^{12}& w^{11}& 0& 0& 0& 0\\
	\end{pmatrix}.
\end{equation}
The matrix \(D_1\), defined by (\ref{eq:imN1}), can be explicitly written as
\begin{equation}
	\label{eq:D1}
	D_1=
	\begin{pmatrix}
		n^1& -n^2& n^3& 0& 0& 0& 0& 0& 0\\
		0& 0& 0& n^1& -n^2& n^3& 0& 0& 0\\
		0& 0& 0& 0& 0& 0& n^1& -n^2& n^3\\
		\rho^1& 0& 0& -\rho^2& 0& 0& \rho^3& 0& 0\\
		0& \rho^1& 0& 0& -\rho^2& 0& 0& \rho^3& 0\\
		0 &	0& \rho^1& 0& 0& -\rho^2& 0& 0& \rho^3
	\end{pmatrix}.
\end{equation}
It is easy to check that \(\rk D_1=5\) at a generic point  (since, for example, its minor corresponding to the columns \(\{1,2,3,6,9\}\) equals to \((n^3)^3(\rho^1)^2\), and does not vanish identically), and that desired equation \(D_1D_0=0\) holds on \(\mathcal{M}\), due to trivial identities
\begin{equation}
	\label{eq:27}
	n^iw^{jk}-n^kw^{ji}=0,\qquad \rho^iw^{kj}-\rho^kw^{ij}=0
\end{equation}
for \(w^{ij}=\rho^in^j\).
Now let us find \(\ker D_1\), namely all vectors  \(D_1\vec{v}=0\) polynomial in \(n^i\) and \(\rho^i\).

Let us now take the first three rows of the matrix \(D_1\) \eqref{eq:D1} and solve the corresponding equations \((D_1\vec{v})^1=(D_1\vec{v})^2=(D_1\vec{v})^3=0\) using Lemma~\ref{thm:lemma1} with \(\{x_1,x_2,x_3\}=\{n^1,n^2,n^3\}\):
\begin{equation}
	\label{eq:vAnsatz}
	\vec{v}=
	\begin{pmatrix}
		c^{13}n^2+c^{12}n^3 \\ c^{13}n^1+c^{11}n^3 \\ c^{11}n^2-c^{12}n^1 \\
		c^{23}n^2+c^{22}n^3 \\ c^{23}n^1+c^{21}n^3 \\ c^{21}n^2-c^{22}n^1 \\
		c^{33}n^2+c^{32}n^3 \\ c^{33}n^1+c^{31}n^3 \\ c^{31}n^2-c^{32}n^1 
	\end{pmatrix}.
\end{equation}
and $c^{ij}=c^{ij}(\vec{n})$ are some polynomials, corresponding to the matrix elements of the matrices $\Omega^{(i)}$, $i=1,2,3$.
Then we rewrite the equations, corresponding to the 4th and 5th rows of \(D_1\) \eqref{eq:D1}
\begin{equation}\begin{gathered}
		\label{eq:30}
		\rho^1(c^{13}n^2+c^{12}n^3)-
		\rho^2(c^{23}n^2+c^{22}n^3)+
		\rho^3 (c^{33}n^2+c^{32}n^3) = 0,\\
		\rho^1(c^{13}n^1+c^{11}n^3)-
		\rho^2(c^{23}n^1+c^{21}n^3)+
		\rho^3(c^{33}n^1+c^{31}n^3)=0.
	\end{gathered}
\end{equation}
as
\begin{equation}\begin{gathered}
		\label{eq:32}
		n^2(c^{13}\rho^1-
		c^{23}\rho^2+
		c^{33}\rho^3)+
		n^3(c^{12}\rho^1-
		c^{22}\rho^2+
		c^{32}\rho^3) = 0,\\
		n^1(c^{13}\rho^1-
		c^{23}\rho^2+
		c^{33}\rho^3)+
		n^3(c^{11}\rho^1-
		c^{21}\rho^2+
		c^{31}\rho^3)=0.
	\end{gathered}
\end{equation}
For the same reason, an obvious consequence of these equations is:
\begin{equation}\begin{gathered}\begin{split}
			\label{eq:34}
			c^{11}\rho^1-c^{21}\rho^2+c^{31}\rho^3&=An^1,\\
			c^{12}\rho^1-c^{22}\rho^2+c^{32}\rho^3&=An^2,\\
			c^{13}\rho^1-c^{23}\rho^2+c^{33}\rho^3&=-An^3.
		\end{split}
	\end{gathered}
\end{equation}
and now one can solve each of these equations for \(A\) and \(c^{ij}\), applying again Lemma~\ref{thm:lemma1} now with \(\{x_1,x_2,x_3,x_4\}=\{\rho^1,\rho^2,\rho^3,n^j\}\):
\begin{equation}\begin{gathered}\begin{split}
			\label{eq:29}
			A&=d^1\rho^1+d^2\rho^2+d^3\rho^3,\\\\
			c^{1j}&=d^{j3}\rho^2+d^{j2}\rho^3+d^1n^j,\\
			c^{2j}&=d^{j3}\rho^1+d^{j1}\rho^3-d^2n^j,\\
			c^{3j}&=-d^{j2}\rho^1+d^{j1}\rho^2+d^3n^j,\\\\
			&j=1,2,3.
		\end{split}
	\end{gathered}
\end{equation}
Substituting this result into \eqref{eq:vAnsatz} one gets
\begin{equation}
	\label{eq:26}
	\vec{v}=
	\begin{pmatrix}
		(d^{33}\rho^2+d^{32}\rho^3-d^1n^3)n^2+(d^{23}\rho^2+d^{22}\rho^3+d^1n^2)n^3\\
		(d^{33}\rho^2+d^{32}\rho^3-d^1n^3)n^1+(d^{13}\rho^2+d^{12}\rho^3+d^1n^1)n^3\\
		(d^{13}\rho^2+d^{12}\rho^3+d^1n^1)n^2-(d^{23}\rho^2+d^{22}\rho^3+d^1n^2)n^1\\
		(d^{33}\rho^1+d^{31}\rho^3+d^2n^3)n^2+(d^{23}\rho^1+d^{21}\rho^3-d^2n^2)n^3\\
		(d^{33}\rho^1+d^{31}\rho^3+d^2n^3)n^1+(d^{13}\rho^1+d^{11}\rho^3-d^2n^1)n^3\\
		(d^{13}\rho^1+d^{11}\rho^3-d^2n^1)n^2-(d^{23}\rho^1+d^{21}\rho^3-d^2n^2)n^1\\
		(-d^{32}\rho^1+d^{31}\rho^2-d^3n^3)n^2+(-d^{22}\rho^1+d^{21}\rho^2+d^3n^2)n^3\\
		(-d^{32}\rho^1+d^{31}\rho^2-d^3n^3)n^1+(-d^{12}\rho^1+d^{11}\rho^2+d^3n^1)n^3\\
		(-d^{12}\rho^1+d^{11}\rho^2+d^3n^1)n^2-(-d^{22}\rho^1+d^{21}\rho^2+d^3n^2)n^1
	\end{pmatrix}.
\end{equation}
After some simplification and substitution \(\rho^in^j=w^{ij}\) we can rewrite it as
\begin{equation}
	\label{eq:28}
	\vec{v}=
	\begin{pmatrix}
		d^{33}w^{22}+d^{32}w^{32}+d^{23}w^{23}+d^{22}w^{33}\\
		d^{33}w^{21}+d^{32}w^{31}+d^{13}w^{23}+d^{12}w^{33}\\
		d^{13}w^{22}+d^{12}w^{32}-d^{23}w^{21}-d^{22}w^{31}\\
		d^{33}w^{12}+d^{31}w^{32}+d^{23}w^{13}+d^{21}w^{33}\\
		d^{33}w^{11}+d^{31}w^{31}+d^{13}w^{13}+d^{11}w^{33}\\
		d^{13}w^{12}+d^{11}w^{32}-d^{23}w^{11}-d^{21}w^{31}\\
		-d^{32}w^{12}+d^{31}w^{22}-d^{22}w^{13}+d^{21}w^{23}\\
		-d^{32}w^{11}+d^{31}w^{21}-d^{12}w^{13}+d^{11}w^{23}\\
		-d^{12}w^{12}+d^{11}w^{22}+d^{22}w^{11}-d^{21}w^{21}
	\end{pmatrix}
	=
	D_0
	\begin{pmatrix}
		d^{11} \\ d^{21} \\ -d^{31} \\ d^{12} \\ d^{22} \\ -d^{32} \\ -d^{13} \\ -d^{23} \\ d^{33}
	\end{pmatrix},
\end{equation}
so this vector lies in the image of \(D_0\), and therefore \(\ker D_1=\im D_0\).

%
%




\end{document}